\documentclass[conference,letterpaper]{IEEEtran}
\IEEEoverridecommandlockouts

\makeatletter
\newcommand\HUGE{\@setfontsize\Huge{40}{50}}
\makeatother

\usepackage[utf8]{inputenc} 
\usepackage[T1]{fontenc}
\usepackage{ifthen}
\usepackage{mathrsfs}
\usepackage{amsthm}
\usepackage{balance}
\usepackage{nicefrac}
\usepackage{multirow}
\usepackage{multicol}
\usepackage[dvipsnames]{xcolor}
\usepackage{tikz}
\usepackage{pgfplots}
\pgfplotsset{compat=1.18}
\usepackage{comment}
\usepackage{stmaryrd}
\usepackage{mathtools}
\usepackage{bm}
\usepackage{amssymb}
\usepackage{graphicx}
\usepackage{textcomp}
\usepackage[caption=false]{subfig}
\usepackage[colorinlistoftodos,textsize=tiny]{todonotes}
\usepackage{booktabs}
\usepackage{stmaryrd}
\usepackage{braket}
\usepackage{nicematrix} %
\usepackage[noadjust]{cite} %

\makeatletter
\renewcommand*\env@matrix[1][*\c@MaxMatrixCols c]{%
  \hskip -\arraycolsep
  \let\@ifnextchar\new@ifnextchar
  \array{#1}}
\makeatother

\usepackage[ruled, vlined, linesnumbered]{algorithm2e}

\SetCommentSty{mycommfont}

\newtheoremstyle{bolddefinition}    %
  {\topsep}                         %
  {\topsep}                         %
  {\itshape}                        %
  {}                                %
  {\bfseries}                       %
  {.}                               %
  { }                               %
  {\thmname{#1}\thmnumber{ #2}\thmnote{ {\normalfont(#3)}}} %

\theoremstyle{bolddefinition}
\newtheorem{proposition}{Proposition}

\newtheorem{theorem}{Theorem}

\theoremstyle{bolddefinition}
\newtheorem{definition}{Definition}

\newtheorem{lemma}{Lemma}

\newtheorem{remark}{Remark}

\newtheorem{example}{Example}

\newcommand{\wt}{\operatorname{wt}}
\newcommand{\rk}{\operatorname{rank}}
\newcommand{\Rspan}{\operatorname{Col}}
\newcommand{\Aut}{\operatorname{Aut}}

\newcommand{\dual}[1]{{#1}^\perp} %
\newcommand{\tn}[1]{\textnormal{#1}}
\newcommand{\vect}[1]{\bm{#1}} %
\newcommand{\Integers}{\mathbb{Z}}   %
\newcommand{\Naturals}{\mathbb{N}}   %
\newcommand{\Field}{\mathbb{F}}      %

\newcommand{\eqdef}{\triangleq} %

\renewcommand{\set}[1]{\mathcal{#1}}           %
\newcommand{\ecard}[1]{|#1|}

\newcommand{\mat}[1]{\mathsf{#1}} %
\newcommand{\trans}[1]{#1^{\textup{\textsf{\tiny T}}}} %

\newcommand{\einner}[2]{\langle{#1},{#2}\rangle}

\newcommand{\code}[1]{\mathscr{#1}} %

\newcommand*{\Scale}[2][4]{\scalebox{#1}{\ensuremath{#2}}} %

\usepackage[acronym,shortcuts]{glossaries}
\newacronym{CSS}{CSS}{Calderbank--Shor--Steane}
\newacronym{DPM}{DPM}{dyadic permutation matrix}
\newacronym{LER}{LER}{logical error rate}
\newacronym{LDPC}{LDPC}{low-density parity-check}
\newacronym{QC}{QC}{quasi-cyclic}
\newacronym{QC-LDPC}{QC-LDPC}{quasi-cyclic low-density parity-check}
\newacronym{QD}{QD}{quasi-dyadic}
\newacronym{QD-LDPC}{QD-LDPC}{quasi-dyadic low-density parity-check}
\newacronym{BP}{BP}{belief propagation}
\newacronym{BDD}{BDD}{bounded distance decoder}
\newacronym{MS}{MS}{min-sum}
\newacronym{BP+OSD}{BP+OSD}{BP plus ordered statistics decoding}
\newacronym{QECC}{QECC}{quantum error correcting code}
\newacronym{QLDPC}{QLDPC}{quantum low-density parity-check}
\newacronym{MSD}{MSD}{magic-state distillation}
\newacronym{GM}{GM}{generator matrix}
\newacronym{GB}{GB}{Generalized bicycle}
\newacronym{ILP}{ILP}{integer linear programming}
\newacronym{LP}{LP}{Lifted-product}
\newacronym{DC}{DC}{dual-containing}
\newacronym{BER}{BER}{bit error rate}
\newacronym{SNR}{SNR}{signal-to-noise ratio}
\newacronym
  [longplural={parity-check matrices},
   shortplural={PCMs}]
  {PCM}{PCM}{parity-check matrix}
\newacronym
  [longplural={quantum stabilizer codes},
   shortplural={QSCs}]
  {QSC}{QSC}{quantum stabilizer code}
\newacronym{QEC}{QEC}{Quantum error correction}
\newacronym{BP4}{BP4}{quaternary belief propagation}
\newacronym{GRAND}{GRAND}{guessing random additive-noise decoding}
\newacronym{aSCED}{aSCED}{affine subcode ensemble decoding}
\newacronym{AED}{AED}{automorphism ensemble decoding}
\newacronym{AED+OSD}{AED$+$OSD}{automorphism ensemble decoding plus ordered statistics decoding}
\newacronym{OSD}{OSD}{ordered statistics decoding}

\pdfmapfile{+custom.map}

\usepackage{amsmath}

\usepackage{array}
\usepackage{enumitem}

\allowdisplaybreaks

\pgfplotsset{
    cycle list={
        {red, mark=*},
        {red, mark=x},
        {red, mark=square*},
        {red, mark=triangle*},  {blue,mark=triangle*},
        {red, mark=diamond*},   {blue,mark=diamond*},
        {red, mark=pentagon*},  {blue,mark=pentagon*}
    },
    legend style={
        at={(0.5,-0.2)},
        anchor=north,
        legend columns=2,
        cells={anchor=west},
        font=\footnotesize,
        rounded corners=2pt,
    },
    xlabel=$r$,
    ylabel=$\text{DFR}$,
}

\newboolean{arXiv} %
\setboolean{arXiv}{true} %

\newlength{\TriPlotWidth}
\ifthenelse{\boolean{arXiv}}{%
  \setlength{\TriPlotWidth}{0.48\textwidth}
}{%
  \setlength{\TriPlotWidth}{0.65\columnwidth}
}

\ifthenelse{\boolean{arXiv}}{%
  \pgfplotsset{tri axis/.style={}, tri cost axis/.style={}}%
  \pgfplotsset{cost noleg/.style={}}%
  \newcommand{\costentry}[1]{\addlegendentry{#1}}%
  \newcommand{\aedentry}[2]{\addlegendentry{#1}}%
}{%
  \pgfplotsset{
    tri base/.style={%
      width=0.260\linewidth, height=5.5cm,
      tick label style={font=\scriptsize},
      label style={font=\footnotesize},
      xlabel style={yshift=1.2mm}, ylabel style={font=\scriptsize, yshift=-0.6mm},
      every axis plot post/.append style={mark size=1.4pt}},
    tri axis/.style={tri base,
      legend to name=#1, legend columns=7,
      legend cell align=center,
      legend style={font=\scriptsize, inner xsep=3pt, inner ysep=2pt, row sep=0pt, fill opacity=1, text opacity=1, /tikz/every even column/.append style={column sep=0.6em}}},
    tri cost axis/.style={tri base,
      legend pos= south east, legend columns=1,
      legend style={font=\tiny, inner sep=1pt, row sep=-1pt, rounded corners=1pt,
                    fill=white, fill opacity=0.85, text opacity=1}},
  }%
  \pgfplotsset{cost noleg/.style={forget plot}}%
  \newcommand{\costentry}[1]{}%
  \newcommand{\aedentry}[2]{\addlegendentry{#2}}%
}

\ifthenelse{\boolean{arXiv}}{%
\usepackage{hyperref}
}{

}

\begin{document}

\title{
On Constructing and Decoding Quantum Triorthogonal Codes
\thanks{This work was supported in part by the Research Council of Norway (RCN) under the NISQEC project (grant no.~$357698$). The work of Alessio Baldelli was partially supported by Agenzia per la Cybersicurezza Nazionale (ACN) under the programme for promotion of XL cycle PhD research in cybersecurity (CUP I32B24001750005).}
}

\author{
\IEEEauthorblockN{
Alessio Baldelli\IEEEauthorrefmark{2},    
Olai \AA.~Mostad\IEEEauthorrefmark{3}\IEEEauthorrefmark{4},
Hsuan-Yin Lin\IEEEauthorrefmark{3},
Eirik Rosnes\IEEEauthorrefmark{3},
Massimo Battaglioni\IEEEauthorrefmark{2}
}
\IEEEauthorblockA{
  \IEEEauthorrefmark{2}\emph{Department of Information Engineering}, Universit{\`a} Politecnica delle Marche, Ancona 60131, Italy
  \\
}
\IEEEauthorblockA{
\IEEEauthorrefmark{3}\emph{Simula UiB}, N--5006 Bergen, Norway%
,\quad
\IEEEauthorrefmark{4}\emph{Department of Informatics}, University of Bergen, N--5020 Bergen, Norway\\
Emails: \{\texttt{a.baldelli@pm.,m.battaglioni@\}univpm.it}%
  , \texttt{\{olai,lin,eirikrosnes\}@simula.no}
}
}

\maketitle

\begin{abstract}
    Triorthogonal codes are a basic ingredient in magic-state distillation, as they enable transversal implementation of logical $\mathsf{T}$ gates. As a non-Clifford operation, the $\mathsf{T}$ gate enables universal fault-tolerant quantum computation when combined with Clifford operations. We derive an existence criterion for even-weight triorthogonal generator matrices with a target dual minimum distance. The criterion combines triorthogonality constraints with MacWilliams identities via Krawtchouk-polynomial conditions on the dual weight distribution, yielding an integer linear programming formulation for the construction problem. We find novel nontrivial triorthogonal codes (degenerate and non-degenerate) that are not necessarily constructed from triply even binary linear codes. By imposing automorphism-group constraints, the formulation also recovers the known $\llbracket 49,1,5\rrbracket$ and $\llbracket 95,1,7\rrbracket$ triorthogonal codes. We also evaluate the performance of the obtained high-distance triorthogonal codes over the dephasing channel using different decoding methods, showing that automorphism ensemble decoding (AED) and guessing random additive-noise decoding (GRAND) are promising options.
\end{abstract}

\section{Introduction}
\label{sec:intro}
Quantum error correction protects quantum information against decoherence and noise.
The stabilizer formalism~\cite{Gottesman97_1} paved the way to several quantum code families, obtained by adapting classical coding theory notions to the quantum setting. 
A notable example is given by \ac{CSS} codes~\cite{CalderbankShor96_1, Steane96_1}.
Nevertheless, not all quantum codes arise as adaptations of generic classical error-correcting codes. An important example is given by \emph{triorthogonal} codes. These quantum CSS codes were introduced in the context of \emph{\ac{MSD}}, a standard approach to implementing costly non-Clifford operators, necessary to achieve universal fault-tolerant quantum computation~\cite{BravyiKitaev05_1, BravyiHaah12_1, HaahHastingsPoulinWecker17_1, HaahHastings18_1, NezamiHaah22_1}.
Triorthogonal codes are especially relevant in this framework because they support transversal $\mat{T}$-gate implementations.
This property is associated with the pairwise and triple-wise overlap constraints that define triorthogonality~\cite{BravyiHaah12_1, NezamiHaah22_1, JainAlbert25_1}. The same structure that makes triorthogonal codes useful also makes their construction challenging, not only because of the aforementioned overlap conditions, but also because of additional row-weight constraints.\looseness-1

This paper studies the construction of binary triorthogonal codes with given dual-distance properties. We derive an existence criterion for even-weight triorthogonal generator matrices by combining the overlap constraints with Krawtchouk-polynomial conditions derived from the MacWilliams identities. This yields an \ac{ILP} formulation that can be used to search for triorthogonal matrices satisfying a target dual minimum distance. The proposed framework connects the algebraic constraints required for transversal $\mat{T}$ gates with distance-oriented design criteria from classical coding theory. This construction problem is related to, but distinct from, previous work on \ac{MSD} and on the coding-theoretic structure of triorthogonal matrices. 
Bravyi and Haah introduced triorthogonal codes as a mechanism for low-overhead distillation of \(\mat{T}\)-type magic states~\cite{BravyiHaah12_1}, presenting instances whose even-weight subcodes are triply even~\cite{BetsumiyaMunemasa12_1}.  
More recently, Shi~\emph{et al.} studied triorthogonal matrices from a classical coding-theoretic perspective, constructing them from binary self-dual codes~\cite{ShiLuKimSole24_1}. 
In contrast, the present work focuses on a complementary existence and design question: given a desired dual-distance target, we ask under which algebraic and enumerative conditions a triorthogonal matrix can be obtained. 
Our ILP formulation yields novel nontrivial triorthogonal codes~(see Table~\ref{tab:code-parameters_nontrivial-TriOrthogonal-codes}), both degenerate and non-degenerate, that are not necessarily generated by triply even codes. The formulation also recovers the known $\llbracket 49,1,5\rrbracket$ and $\llbracket 95,1,7\rrbracket$ codes when suitable automorphism-group constraints are imposed.

\ifthenelse{\boolean{arXiv}}{%
Finally, we evaluate the performance of triorthogonal codes over the dephasing channel, which is the relevant noise model for the \ac{MSD} protocol, since only phase-flip errors need to be considered~\cite{BravyiHaah12_1}.
The constructions considered in this work also provide explicit permutation automorphisms that can be used for decoding. For the new $\llbracket89,1,5\rrbracket$ code found by our \ac{ILP} formulation, we exploit the permutation automorphisms arising from its cyclic structure. For the $\llbracket95,1,7\rrbracket$ code of~\cite{Sullivan24_1, JainAlbert25_1}, we exploit the permutation automorphisms inherited from the constituent codes, together with the additional symmetries provided by the doubling construction. These symmetries give a systematic way to generate alternative \ac{BP} decoding paths, motivating the use of \ac{AED}, which has been studied for classical~\cite{KriegClausiusRuebenacketenBrink25_1, MandelbaumBeznerJakeltenBrinkSchmalen26_1sub} and quantum~\cite{KoutsioumpasSayginelWebsterBrowne25_1sub} codes. For the considered nonsparse \acp{PCM}, our numerical results show that combining these paths with \ac{OSD} post-processing substantially improves the decoding performance. We compare this approach with \ac{GRAND}~\cite{DuffyLiMedard19_1}, which was previously used for random \acp{QSC} in~\cite{CruzMonteiroCoutinho23_1}. The comparison illustrates how the symmetries inherited from the code construction can be used to obtain a good trade-off between decoding performance and complexity.
}{%
Finally, we evaluate the obtained codes over the dephasing channel, which is the relevant noise model for \ac{MSD}~\cite{BravyiHaah12_1}. The permutation automorphisms inherited from their constructions, namely the cyclic structure of our new $\llbracket 89,1,5\rrbracket$ code and the doubling structure of the $\llbracket 95,1,7\rrbracket$ code~\cite{Sullivan24_1, JainAlbert25_1}, provide alternative \ac{BP} decoding paths for \ac{AED}~\cite{KriegClausiusRuebenacketenBrink25_1, MandelbaumBeznerJakeltenBrinkSchmalen26_1sub, KoutsioumpasSayginelWebsterBrowne25_1sub}. Combined with \ac{OSD} post-processing, these paths substantially improve the decoding of nonsparse \acp{PCM}, and we compare this approach with \ac{GRAND}~\cite{DuffyLiMedard19_1, CruzMonteiroCoutinho23_1}.
}

\section{Notation and Preliminaries}
\label{sec:preliminaries}

In this paper, $\Field_2$ denotes the binary field, $\Integers$ the set of integers, $\Naturals_{0} \eqdef \{0,1,2,\ldots\}$  the set of nonnegative integers, and $\mathbb{C}$ the set of complex numbers. For a positive integer $n$, we use the shorthand $[n]\eqdef\{1,2,\ldots,n\}$. Column vectors, e.g., $\vect{a}$, and matrices, e.g., $\mat{A}$, are denoted by bold lowercase letters and sans-serif uppercase letters, respectively. For a vector $\vect{a}$, its $i$-th entry is denoted by $a_i$ or $[\vect{a}]_i$. Similarly, for a matrix $\mat{A}$, the entry in the $i$-th row and $j$-th column is denoted by $[\mat{A}]_{i,j}$. The transpose of a vector (matrix) is denoted as $\trans{\bm{a}}$ ($\trans{\mat{A}}$). All-zero vectors and matrices are denoted by $\bm{0}$ and $\mat 0$, respectively.
Unless otherwise specified, operations involving binary vectors and binary matrices are performed over $\Field_2$.
The inner product between two vectors $\vect{a}$ and $\vect{b}$ is denoted by $\einner{\vect{a}}{\vect{b}}$, and the  Hamming weight (or, simply, weight) of a vector $\vect{a}$, i.e., the number of nonzero elements of $\vect{a}$, is denoted by $\wt(\bm{a})$. 
The rank of a binary matrix $\mat{A}$ is denoted by $\rk(\mat{A})$, while its column span is denoted by $\Rspan(\mat{A})$.

\subsection{Classical and Quantum Codes}
\label{sec:classical-quantum-codes}

A classical binary linear code of length $n$, dimension $k$, and minimum Hamming distance $d$ is denoted by $[n,k,d]$ and is a subspace $\code{C}\subseteq \mathbb{F}_2^{n}$.  
The vectors $\vect{c} \in \code{C}$ are called \emph{codewords}, and  
the minimum Hamming distance $d$ of $\code{C}$ is the minimum weight of its nonzero codeword(s). 
Then, its error-correcting capability is
    $t = \left\lfloor\nicefrac{(d-1)}{2} \right\rfloor$.
A code can be represented as the kernel of a \ac{PCM} $\mat{H}\in \mathbb{F}_2^{m \times n}$ or, equivalently, through a (full-rank) \ac{GM} $\mat{G} \in \mathbb{F}_2^{k \times n}$, namely $\code{C} \eqdef \left\{\bm{c}\in\mathbb F_2^n \mid \mat{H}\bm{c} = \bm{0}\right\} =\left\{\trans{\mat{G}}\bm{m}\mid \bm{m}\in\mathbb F_2^k\right\}$.
The redundancy of $\code{C}$ is $r\eqdef n-k=\rk(\mat{H})\leq m$. We denote by $\code{C}^{\perp} \subseteq \mathbb{F}_2^{n}$ the  $[n,n-k,\dual{d}]$ \emph{dual code} of $\code{C}$, defined as \(
    \code{C}^{\perp}
    \eqdef
    \left\{
    \bm{x}\in\mathbb{F}_2^n \mid
    \einner{\bm{x}}{\bm{c}}=0, \, \forall \, \bm{c}\in\code{C}
    \right\}.
\) Then,
\(
    \dual{\code{C}}=\Rspan(\trans{\mat{H}}).
\)
A code $\code{C}$ satisfying $\code{C}\subseteq\code{C}^{\perp}$ is called \emph{self-orthogonal}.
A binary classical code is called \emph{triply even} if every codeword has Hamming weight divisible by $8$. \ifthenelse{\boolean{arXiv}}{%
An \emph{action} of a group $\Gamma$ on a nonempty set $\set{S}$ is a mapping $\Gamma\times\set{S}\to\set{S}$, $(\gamma,s)\mapsto\gamma(s)$, such that $(\gamma\gamma')(s)=\gamma(\gamma'(s))$ and $\operatorname{id}(s)=s$ for all $\gamma,\gamma'\in\Gamma$ and $s\in\set{S}$, where $\operatorname{id}$ denotes the identity of $\Gamma$. The set $\Gamma(s)\eqdef\{\gamma(s)\mid\gamma\in\Gamma\}$ is called the \emph{$\Gamma$-orbit} of $s$, and the $\Gamma$-orbits form a partition of $\set{S}$. A permutation $\pi$ of the $n$ coordinates is a \emph{permutation automorphism} of a code $\code{C}\subseteq\Field_2^n$ if $\pi(\code{C})=\code{C}$; the permutation automorphisms of $\code{C}$ form a group, denoted by $\Aut(\code{C})$.
}{}

An $\llbracket n,k,d \rrbracket$ \ac{QSC} is a $2^k$-dimensional subspace $\code{C}\subseteq(\mathbb{C}^2)^{\otimes n}$ with minimum distance $d$~\cite{Gottesman97_1}. %
In this paper, we focus on \ac{CSS} codes, i.e., \acp{QSC} whose stabilizer generators are either pure $\mat{X}$-type or pure $\mat{Z}$-type operators~\cite{CalderbankShor96_1, Steane96_1, CalderbankRainsShorSloane98_1}.

\begin{definition}[\ac{CSS} code]
  \label{def:CSS_code}
  An $\llbracket n,k,d \rrbracket$ quantum \ac{CSS} code $\code{C}$ is defined by two classical binary codes $\code{C}_\tn{X}$ and $\code{C}_\tn{Z}$ with parameters $[n,k_\tn{X},d_0]$ and $[n,k_\tn{Z},d_1]$, respectively, represented by $\mat{H}_\tn{X} \in \mathbb{F}_2^{m_\tn{X} \times n}$ and $\mat{H}_\tn{Z}\in \mathbb{F}_2^{m_\tn{Z} \times n}$, and satisfying $\code{C}_\tn{X}^{\perp} \subseteq \code{C}_\tn{Z}$.
  This is equivalent to the orthogonality condition
  $\mat{H}_\tn{X}\trans{\mat{H}}_\tn{Z} = \mat{0}$.
\end{definition}
The dimension of the \ac{CSS} code $\code{C}$ turns out to be $k = k_\tn{X} + k_\tn{Z}-n$. 
The structure of \Ac{CSS} codes allows us to carry out the correction of $\mat{Z}$-type errors and $\mat{X}$-type errors independently at the decoder level.
In particular, the classical code $\code{C}_{\tn{X}}$ is used to correct $\mat{Z}$-type errors, whereas the other classical code, $\code{C}_{\tn{Z}}$, is employed to correct $\mat{X}$-type errors. 
It is important to mention that the (quantum) minimum distance of the $\llbracket n, k, d \rrbracket$ \ac{CSS} code $\code{C}$ is computed as 
 $d \eqdef \min\{d_{\tn{X}}, d_{\tn{Z}}\}$,
where
\begin{IEEEeqnarray*}{rCl}
    d_{\tn{X}} & \eqdef &\min\{ \wt(\bm{u}) \, | \, \bm{u} \in \code{C}_{\tn{X}} \backslash \code{C}_{\tn{Z}}^{\perp} \} \geq {d_0}, 
    \\[-1mm]
    d_{\tn{Z}} & \eqdef &\min\{ \wt(\bm{v}) \, | \, \bm{v} \in \code{C}_{\tn{Z}} \backslash \code{C}_{\tn{X}}^{\perp} \} \geq {d_1}.
\end{IEEEeqnarray*}
It follows that $d \geq \min\{ d_0, d_1 \}$. If $d > \min\{ d_0, d_1 \}$, the \ac{CSS} code $\code{C}$ is called \emph{degenerate}, and if $d_{\tn{X}} > d_0$ ($d_{\tn{Z}} > d_1$), it is called \emph{$\mat{X}$-degenerate} (\emph{$\mat{Z}$-degenerate}).
The  capability of correcting $\mat{Z}$-type and $\mat{X}$-type errors is $t_{\tn{X}} = \left\lfloor\nicefrac{(d_{\tn{X}}-1)}{2} \right\rfloor$ %
and $t_{\tn{Z}} = \left\lfloor\nicefrac{(d_{\tn{Z}}-1)}{2} \right\rfloor$, respectively.

\subsection{Noise Model} 
\label{subsec:error_model}
We focus on the dephasing error model, where each qubit undergoes a $\mat{Z}$-type, or phase-flip, error with \emph{dephasing probability} $p$. Equivalently, the error pattern is generated by a binary symmetric channel. For \ac{CSS} codes, this allows us to use a binary decoder based only on the $\mat{X}$-type stabilizer checks, represented by $\mat{H}_{\tn{X}}\in\mathbb{F}_2^{m_{\tn{X}}\times n}$. After syndrome measurement, the decoder receives $\bm{s}_{\tn{X}}=\mat{H}_{\tn{X}}\bm{e}_{\tn{Z}}\in\mathbb{F}_2^{m_{\tn{X}}}$, where $\bm{e}_{\tn{Z}}\in\mathbb{F}_2^n$ identifies the qubit positions affected by phase-flip errors.

\subsection{Triorthogonal Codes} 
\label{subsec:tri_codes}

We first recall some basic facts regarding triorthogonal matrices and codes, both in the classical and quantum settings~\cite{BravyiHaah12_1}.
\begin{definition}[Triorthogonal matrix]
  \label{def:TriOrthogonal-GM}
  A matrix $\mat{A} \in \mathbb{F}_2^{r \times n}$ is called \emph{triorthogonal} if it satisfies
  \begin{equation*}
    \sum_{j = 1}^{n} [\mat{A}]_{a, j} [\mat{A}]_{b, j} = 0\quad \text{and}\quad\sum_{j = 1}^{n} [\mat{A}]_{a, j} [\mat{A}]_{b, j} [\mat{A}]_{c, j} = 0
  \end{equation*}%
  for all pairs $1 \leq a < b \leq r$ and for all triples $1 \leq a < b < c \leq r$, respectively.
\end{definition}

\begin{remark}
  \label{rem:even-overlap_Hx}
  A matrix $\mat{A} \in \mathbb{F}_2^{r \times n}$ is triorthogonal if and only if \emph{the supports of any pair} and \emph{any triple of its rows have even overlap}. %
\end{remark}

\begin{definition}[Classical \ac{GM} to construct a triorthogonal code]
  \label{def:G_CSS_code}
  Let 
  \begin{equation*} 
    \label{eq:G_triorto}
    \mat{G}_{\tn{Z}} = 
    \begin{pmatrix}
      \mat{G}_1
      \\
      \mat{H}_{\tn{X}}
    \end{pmatrix} \in \mathbb{F}_2^{(k + k_0 ) \times n}
  \end{equation*}
  be a triorthogonal \ac{GM} where $\mat{G}_1 \in \mathbb{F}_2^{k \times n}$ has odd-weight rows and $\mat{H}_{\tn{X}} \in \mathbb{F}_2^{k_0 \times n}$ has even-weight rows. We write $\code{C}_{\tn{Z}} \triangleq\Rspan(\trans{\mat{G}_{\tn{Z}}})$, $\code{C}_\tn{X}^{\perp} \triangleq \Rspan(\trans{\mat{H}_{\tn{X}}})$, and $\code{G}_1 \triangleq \Rspan(\trans{\mat{G}_1})$, and the CSS code given by $\code{C}_{\tn{X}}^{\perp} \subseteq \code{C}_{\tn{Z}}$ is called a triorthogonal code.
\end{definition}

\begin{lemma}[\protect{\cite[Lemma 1]{BravyiHaah12_1}}]
  \label{lem:triorto}
  Let $\mat{G}_{\tn{Z}} \in \mathbb{F}_2^{(k+k_0) \times n}$ be triorthogonal. Then, $\mat{G}_1$ is full-rank, $\code{C}_\tn{X}^{\perp} \cap \code{G}_1 = \{\bm 0 \}$, $\code{C}_{\tn{X}}^{\perp} = \code{C}_{\tn{Z}} \cap \code{C}_{\tn{Z}}^{\perp}$, and $\code{C}_\tn{X} = \code{G}_1 \oplus \code{C}_\tn{Z}^{\perp}$.
\end{lemma}

Next, we particularize Definition~\ref{def:CSS_code} for a triorthogonal code.
\begin{proposition}[Triorthogonal code]
  \label{prop:CSS_T_codes}
  Let $\mat{G}_{\tn{Z}} \in \mathbb{F}_2^{(k+k_0) \times n}$ be a triorthogonal GM,  as in Definition~\ref{def:G_CSS_code}. Then, the corresponding \ac{CSS} code of length $n$ has dimension $k$, which is the number of \emph{odd-weight} rows of $\mat{G}_{\tn{Z}}$.
\end{proposition}

\section{Construction of Even-Weight Triorthogonal Matrices with Odd Lengths}
\label{sec:main_result}

In this section, we first construct suitable even-weight triorthogonal matrices $\mat{H}_{\tn{X}}$ of odd length $n$. Then, we append an all-ones row of length $n$ to obtain  $\llbracket n, 1, d \rrbracket$ triorthogonal codes. This follows the standard approach in~\cite{BravyiHaah12_1, JainAlbert25_1}. 

\subsection{Weight Enumerators and MacWilliams Theorem}
\label{sec:weight-enumerators_MacWilliams-Theorem}

For a binary linear $[n,k, d]$ code $\code{C}$, we denote by $A_j$ the number of codewords of weight $j$ in $\code{C}$, i.e., $A_j\eqdef\ecard{\{\vect{c}\in \code{C}\, | \,\wt(\vect{c})=j\}}$, for $j\in \{0\}\cup [n]$. Thus, the \emph{weight enumerator} of $\code{C}$ is given by $W_{\code{C}}(y)=\sum_{w=0}^{n}A_{w} y^{w}$. Similarly, we denote by $B_{\ell}$ the number of codewords of weight $\ell\in \{0\}\cup [n]$ in the dual code $\dual{\code{C}}$. The relationship between the coefficients $A_{j}$ and $B_{\ell}$ is characterized by the MacWilliams theorem.
\begin{theorem}[{MacWilliams theorem~\cite[Ch.~5, Eq.~(13)]{MacWilliamsSloane77_1}}]
  \label{thm:macwilliams-Theorem}
  Let $\code{C}$ be a binary linear code, and let $\dual{\code{C}}$ denote its dual. Then, for every $\ell\in\{0\}\cup [n]$,
  \begin{equation}
    B_{\ell}=\frac{1}{\ecard{\code{C}}}\sum_{j=0}^{n} A_j \mathrm{K}_{\ell}(j;n),
    \label{eq:MacWilliams-Theorem}
  \end{equation}
  where $\mathrm{K}_{\ell}(j;n)$ is the binary Krawtchouk polynomial:
  \begin{equation*}
    \mathrm{K}_{\ell}(j;n)\eqdef\sum_{s=0}^{\ell}(-1)^s\binom{j}{s}\binom{n-j}{\ell-s}.
  \end{equation*}
\end{theorem}

\subsection{Integer Linear Programming  Formulation for Triorthogonal Codes}
\label{sec:ILP}

Inspired by the construction of self-orthogonal codes in~\cite{KohnertWassermann09_1}, we propose a general construction of binary linear codes whose \acp{GM} are triorthogonal and have only even-weight rows, while imposing the dual-distance constraint $d(\dual{\code{C}})\geq \dual{d}$. Before presenting the ILP formulation, we introduce the following notation. Define $\set{V}\eqdef\Field_2^{k_0}\setminus\{\vect{0}\}$ as the set of all nonzero binary column types, and let $N\eqdef\ecard{\set{V}}=2^{k_0}-1$. We enumerate the elements $\vect{v}_1,\ldots,\vect{v}_N\in\set{V}$. For each $\vect{v}_i\in\set{V}$, the variable $x_i\in\Naturals_0$ denotes the multiplicity of the column type $\vect{v}_i$ in the GM. Thus, the \emph{multiplicity vector} $\vect{x}=\trans{(x_1,\ldots,x_N)}$ specifies the multiset of columns of the desired GM.

We next define three matrices that encode the constraints in our formulation. First, let $\mat{M}\in\{0,1\}^{N\times N}$ be the incidence matrix indexed by the nonzero vectors in $\set{V}$, such that
\begin{equation}
  [\mat{M}]_{i,i'}=
  \begin{cases}
    1, & \tn{if } \einner{\vect{v}_i}{\vect{v}_{i'}}=0,
    \\
    0, & \tn{otherwise}.\label{eq:Mk_weights_codewords}
  \end{cases}
\end{equation}
This matrix is used to compute the weights of the nonzero codewords generated by the chosen column multiplicities.

Second, let $\mat{P}_{k_0}\in\{0,1\}^{\binom{k_0+1}{2}\times N}$ be the self-orthogonality constraint matrix. Indexing its rows by pairs $(a,b)$ with $1\leq a\leq b\leq k_0$, its entries are defined by
\begin{equation}
  [\mat{P}_{k_0}]_{(a,b),i}=[\vect{v}_i]_a[\vect{v}_i]_b.\label{eq:Pk_self-orthogonality}
\end{equation}
Therefore, for a given multiplicity vector, the entry $[\mat{P}_{k_0}\vect{x}]_{(a,b)}$ counts the weight of row $a$ when $a=b$, and the pairwise overlap between rows $a$ and $b$ when $a<b$.

Similarly, we define the triorthogonality constraint matrix $\mat{T}_{k_0}\in\{0,1\}^{\binom{k_0}{3}\times N}$. Indexing its rows by triples $(a,b,c)$ with $1\leq a< b< c \leq k_0$, its entries are defined by
\begin{equation}
  [\mat{T}_{k_0}]_{(a,b,c),i}=[\vect{v}_i]_a[\vect{v}_i]_b[\vect{v}_i]_c.\label{eq:Tk_tri-orthogonality}
\end{equation}
Therefore, $[\mat{T}_{k_0}\vect{x}]_{(a,b,c)}$ counts the triple overlap among rows $a$, $b$, and $c$. These matrices allow the self-orthogonality, triorthogonality, and dual-distance requirements to be expressed as linear constraints in the ILP. \ifthenelse{\boolean{arXiv}}{Here, the matrix-vector products and the constraints in the ILP are evaluated over the integers, whereas the inner products defining $\mat{M}$ are evaluated over $\Field_2$.}{}

\begin{theorem}[Triorthogonal codes with dual-distance constraint]
  \label{thm:construction_TriOrthogonal_codes_dperp}
  For $\dual{d}\geq 2$, there exists a binary $[n, {k_0}]$ code $\code{C}$ with $d(\dual{\code{C}}) \geq \dual{d}$ and a full-rank triorthogonal matrix $\mat{H}_{\tn{X}}\in\Field_2^{k_0\times n}$ of $\code{C}$ having only even-weight rows if and only if there exist vectors $\vect{x}\in(\{0\}\cup [n])^{N}$, $\vect{z}_\tn{P}\in\Naturals_{0}^{\binom{{k_0}+1}{2}}, \vect{z}_\tn{T}\in\Naturals_{0}^{\binom{{k_0}}{3}}$ and binary indicators $\delta_{i,j}\in \{0, 1\}$, $i\in [N]$, $j\in [n]$, satisfying
  \begin{IEEEeqnarray*}{rCl l}
    \sum_i x_i& = & n
    & \IEEEeqnarraynumspace\mbox{\tn{(L1)}}\label{eq:length}
    \\
    \mat{P}_{{k_0}}\vect{x} - 2\vect{z}_{\tn{P}}& = & 0, \, \tn{(self-orthogonality)}
    & \IEEEeqnarraynumspace\mbox{\tn{(O1)}}\label{eq:SO}
    \\
    \mat{T}_{k_0}\vect{x} - 2\vect{z}_{\tn{T}} & = & 0, \,\tn{(triorthogonality)}
    & \IEEEeqnarraynumspace\mbox{\tn{(O2)}}\label{eq:tri}
    \\[1mm]
    \sum_{j} \delta_{i,j}& = & 1,\, \forall\, i\in [N], 
    & \IEEEeqnarraynumspace\mbox{\tn{(W1)}}\label{eq:onehot}
    \\[-1mm]
    \sum_j j\,\delta_{i,j} + [\mat{M}\vect{x}]_i & = & n,\,\forall\, i\in [N],
    & \IEEEeqnarraynumspace\mbox{\tn{(W2)}}\label{eq:weight}
    \\[-1mm]
    \sum_{i,j} \mathrm{K}_\ell(j;n)\,\delta_{i,j}
    & = & -\binom{n}{\ell},\, \ell = 1,\ldots,\dual{d}-1,
    & \IEEEeqnarraynumspace\mbox{\tn{(D=)}}\label{eq:Dzero}
  \end{IEEEeqnarray*}
  where $[\mat{M}\vect{x}]_i\eqdef\sum_{i'}[\mat{M}]_{i,i'}x_{i'}$.
\end{theorem}

\ifthenelse{\boolean{arXiv}}{The proof is given in Appendix~\ref{sec:proof_ILP-theorem}. The nonnegativity conditions $B_\ell\geq 0$ for $\ell\geq\dual{d}$, used in the linear-programming bound~\cite[Ch.~17, Th.~18]{MacWilliamsSloane77_1}, are not needed: for any integer solution, $(\tn{W1})$ and $(\tn{W2})$ force the indicators $\delta_{i,j}$ to describe the weight distribution of the code generated by $\mat{H}_{\tn{X}}$, so these conditions hold automatically by~\eqref{eq:MacWilliams-Theorem}. In our experiments, adding them as constraints left the feasible set unchanged but increased the solving time.}{The proof is given in~\cite[Appendix~A]{BaldelliMostadLinRosnesBattaglioni26_1sub}.} Theorem~\ref{thm:construction_TriOrthogonal_codes_dperp} provides an ILP formulation for deciding the existence of a triorthogonal matrix $\mat{H}_{\tn{X}}$ satisfying the target dual-distance constraint, and for constructing one whenever the formulation is feasible. Here, $\code{C}=\dual{\code{C}}_{\tn{X}}$ and $d(\dual{\code{C}})=d_0$. We illustrate the formulation by showing how it reconstructs the $\llbracket 15,1,3 \rrbracket$ triorthogonal code of~\cite{BravyiHaah12_1}. Note that, for the $\llbracket 15,1,3\rrbracket$ and $\llbracket 49,1,5\rrbracket$ triorthogonal codes presented in~\cite{BravyiHaah12_1}, the code generated by $\mat{H}_{\tn{X}}$ is triply even.

\ifthenelse{\boolean{arXiv}}{%
\begin{example}
  \label{ex:ex_triorthogonal_n15k1}
  Consider the case $k_0=4$ and $n=15$, where the columns of $\mat{H}_{\tn{X}}$ are given by all nonzero vectors of $\Field_2^4$, i.e.,
  \begin{IEEEeqnarray*}{c}
    \mat{H}_{\tn{X}}=
    \Scale[0.85]{
      \begin{pmatrix}
        0 & 0 & 0 & 0 & 0 & 0 & 0 & 1 & 1 & 1 & 1 & 1 & 1 & 1 & 1
        \\
        0 & 0 & 0 & 1 & 1 & 1 & 1 & 0 & 0 & 0 & 0 & 1 & 1 & 1 & 1 
        \\
        0 & 1 & 1 & 0 & 0 & 1 & 1 & 0 & 0 & 1 & 1 & 0 & 0 & 1 & 1 
        \\
        1 & 0 & 1 & 0 & 1 & 0 & 1 & 0 & 1 & 0 & 1 & 0 & 1 & 0 & 1
      \end{pmatrix}}. 
  \end{IEEEeqnarray*}
  Here, $\mat{H}_{\tn{X}}$ has only even-weight rows. In fact, it is constant-weight with row weight $8$, and $\set{V}=\Field_2^4\setminus\{\mathbf{0}\}$ and $N=2^4-1=15$. In this example, each column type appears exactly once, so $\vect{x}=\trans{(1,1,\ldots,1)}\in (\{0\}\cup [15])^{N}$. Hence, $\sum_{i=1}^{15}x_i=15=n$, which verifies $(\tn{L1})$.
  
  We illustrate the matrix $\mat{P}_4$ defined in~\eqref{eq:Pk_self-orthogonality}, which records all row weights and pairwise row overlaps as follows: %
  \begin{IEEEeqnarray*}{c}
    \Scale[0.5]{
      \setlength{\arraycolsep}{0.0pt}
      \begin{NiceArray}{c*{15}{c}}[first-row,first-col]
        & \vect{v}_1 & \vect{v}_2 & \vect{v}_3 & \vect{v}_4 & \vect{v}_5
        & \vect{v}_6 & \vect{v}_7 & \vect{v}_8 & \vect{v}_9 & \vect{v}_{10}
        & \vect{v}_{11} & \vect{v}_{12} & \vect{v}_{13} & \vect{v}_{14} & \vect{v}_{15}
        \\ \hline
        & \trans{(0001)} & \trans{(0010)} & \trans{(0011)}
        & \trans{(0100)} & \trans{(0101)} & \trans{(0110)}
        & \trans{(0111)} & \trans{(1000)} & \trans{(1001)}
        & \trans{(1010)} & \trans{(1011)} & \trans{(1100)}
        & \trans{(1101)} & \trans{(1110)} & \trans{(1111)}
        \\ \hline
        (1,1) & 0&0&0&0&0&0&0&1&1&1&1&1&1&1&1
        \\
        (1,2) & 0&0&0&0&0&0&0&0&0&0&0&1&1&1&1
        \\
        (1,3) & 0&0&0&0&0&0&0&0&0&1&1&0&0&1&1
        \\
        (1,4) & 0&0&0&0&0&0&0&0&1&0&1&0&1&0&1
        \\
        (2,2) & 0&0&0&1&1&1&1&0&0&0&0&1&1&1&1
        \\
        (2,3) & 0&0&0&0&0&1&1&0&0&0&0&0&0&1&1
        \\
        (2,4) & 0&0&0&0&1&0&1&0&0&0&0&0&1&0&1
        \\
        (3,3) & 0&1&1&0&0&1&1&0&0&1&1&0&0&1&1
        \\
        (3,4) & 0&0&1&0&0&0&1&0&0&0&1&0&0&0&1
        \\
        (4,4) & 1&0&1&0&1&0&1&0&1&0&1&0&1&0&1
      \end{NiceArray}}
    \label{eq:illustration_build_Pk}
  \end{IEEEeqnarray*}
  Note that the rows of $\mat{P}_4$ are ordered as $(1,1)$, $(1,2)$, $(1,3)$, $(1,4)$, $(2,2)$, $(2,3)$, $(2,4)$, $(3,3)$, $(3,4)$, $(4,4)$. Since each row of $\mat{H}_{\tn{X}}$ has weight $8$ and each pair of distinct rows overlaps in $4$ positions, we obtain $\mat{P}_4\vect{x}=\trans{(8,4,4,4,8,4,4,8,4,8)}=2\trans{(4,2,2,2,4,2,2,4,2,4)}$. Hence, $(\tn{O1})$ holds with $\vect{z}_{\tn{P}}=\trans{(4,2,2,2,4,2,2,4,2,4)}$. Similarly, $\mat{T}_4$ defined in~\eqref{eq:Tk_tri-orthogonality} records all triple row overlaps. Since every triple of rows overlaps in exactly $2$ positions, we have $\mat{T}_4\vect{x}=\trans{(2,2,2,2)}=2\trans{(1,1,1,1)}$. Thus, $(\tn{O2})$ holds with $\vect{z}_{\tn{T}}=\trans{(1,1,1,1)}$.
  
  Recall that the rows and columns of $\mat{M}$, defined in~\eqref{eq:Mk_weights_codewords}, are indexed by the nonzero vectors in $\set{V}$. Let $\vect{v}_i\in\set{V}$, such that $\trans{\mat{H}_{\tn{X}}}\vect{v}_i$ is a linear combination of the rows of $\mat{H}_{\tn{X}}$, and let $\vect{v}_{i'}\in\set{V}$ represent a column type of $\mat{H}_{\tn{X}}$. If $\einner{\vect{v}_i}{\vect{v}_{i'}} = 0$, then every column of type $\vect{v}_{i'}$ contributes a zero entry to the codeword $\trans{\mat{H}_{\tn{X}}}\vect{v}_i$. Since $x_{i'}$ is the multiplicity of the column type $\vect{v}_{i'}$, the quantity $[\mat{M}\vect{x}]_i=\sum_{i'=1}^{N}[\mat{M}]_{i,i'}x_{i'}$ counts the number of zero positions in $\trans{\mat{H}_{\tn{X}}}\vect{v}_i$. Consequently, $\wt(\trans{\mat{H}_{\tn{X}}}\vect{v}_i)=n-[\mat{M}\vect{x}]_i$. 
  
  The indicator variables $\delta_{i,j}$ encode the weights of the nonzero codewords $\trans{\mat{H}_{\tn{X}}}\vect{v}_i$, i.e., $\delta_{i,j}=1$ if and only if $\wt(\trans{\mat{H}_{\tn{X}}}\vect{v}_i)=j$. Therefore, the weight enumerator coefficients $\{A_j\}_{j=0}^{n}$ of the code generated by $\mat{H}_{\tn{X}}$ satisfy $A_0=1$ and $A_j=\sum_{i=1}^{N}\delta_{i,j}$ for $1\leq j\leq n$. In this example, every nonzero codeword has weight $8$, so $\delta_{i,8}=1$ for all $i\in[15]$ and $\delta_{i,j}=0$ for $j\neq 8$. Hence, $A_0=1$, $A_8=15$, and $A_j=0$ for all $j\notin\{0,8\}$. This verifies $(\tn{W1})$ and $(\tn{W2})$.  
  
  Finally, the constraints $(\tn{D=})$ are derived from~\eqref{eq:MacWilliams-Theorem} and impose the desired dual-distance condition. For instance, if $\dual{d}=3$, then $(\tn{D=})$ forces the dual code to have no nonzero codewords of weights $1$ and $2$. In this example, $\mathrm{K}_1(8;15)=-1$, $\mathrm{K}_2(8;15)=-7$, and $\mathrm{K}_3(8;15)=7$. Since $A_8=15$ is the only nonzero coefficient with $j\geq 1$, the left-hand side of $(\tn{D=})$ equals $15\,\mathrm{K}_1(8;15)=-15=-\binom{15}{1}$ for $\ell=1$ and $15\,\mathrm{K}_2(8;15)=-105=-\binom{15}{2}$ for $\ell=2$, i.e., $B_1=B_2=0$, and~\eqref{eq:MacWilliams-Theorem} gives $B_3=\nicefrac{(\binom{15}{3}+15\,\mathrm{K}_3(8;15))}{16}=\nicefrac{(455+105)}{16}=35$. Hence, $d_0=3$. Since these weight-$3$ dual codewords have odd weight, they are logical $\mat{Z}$ operators, so $d_{\tn{X}}=3$; together with $d_{\tn{Z}}=15-8=7$, the quantum distance is $3$. Thus, the \ac{ILP} constraints certify that $\mat{H}_{\tn{X}}$ is triorthogonal and that the code generated by $\mat{H}_{\tn{X}}$ satisfies the prescribed dual-distance condition. Appending the all-ones row to this even-weight triorthogonal matrix gives the standard triorthogonal matrix associated with the $\llbracket 15,1,3\rrbracket$ triorthogonal code.  
\end{example}
}{
\begin{example}
  \label{ex:ex_triorthogonal_n15k1}
  Let $k_0=4$ and $n=15$, and let the columns of $\mat{H}_{\tn{X}}$ be the $15$ nonzero vectors of $\Field_2^4$,
  \begin{IEEEeqnarray*}{c}
    \mat{H}_{\tn{X}}=
    \Scale[0.85]{
      \begin{pmatrix}
        0 & 0 & 0 & 0 & 0 & 0 & 0 & 1 & 1 & 1 & 1 & 1 & 1 & 1 & 1
        \\
        0 & 0 & 0 & 1 & 1 & 1 & 1 & 0 & 0 & 0 & 0 & 1 & 1 & 1 & 1
        \\
        0 & 1 & 1 & 0 & 0 & 1 & 1 & 0 & 0 & 1 & 1 & 0 & 0 & 1 & 1
        \\
        1 & 0 & 1 & 0 & 1 & 0 & 1 & 0 & 1 & 0 & 1 & 0 & 1 & 0 & 1
      \end{pmatrix}},
  \end{IEEEeqnarray*}
  so that $\vect{x}=\trans{(1,\ldots,1)}$ and $(\tn{L1})$ holds. The matrix $\mat{P}_4$ in~\eqref{eq:Pk_self-orthogonality}, with rows indexed by the pairs $(a,b)$ and columns by the column types $\vect{v}_i$, is
  \begin{IEEEeqnarray*}{c}
    \Scale[0.5]{
      \setlength{\arraycolsep}{0.0pt}
      \begin{NiceArray}{c*{15}{c}}[first-row,first-col]
        & \vect{v}_1 & \vect{v}_2 & \vect{v}_3 & \vect{v}_4 & \vect{v}_5
        & \vect{v}_6 & \vect{v}_7 & \vect{v}_8 & \vect{v}_9 & \vect{v}_{10}
        & \vect{v}_{11} & \vect{v}_{12} & \vect{v}_{13} & \vect{v}_{14} & \vect{v}_{15}
        \\ \hline
        & \trans{(0001)} & \trans{(0010)} & \trans{(0011)}
        & \trans{(0100)} & \trans{(0101)} & \trans{(0110)}
        & \trans{(0111)} & \trans{(1000)} & \trans{(1001)}
        & \trans{(1010)} & \trans{(1011)} & \trans{(1100)}
        & \trans{(1101)} & \trans{(1110)} & \trans{(1111)}
        \\ \hline
        (1,1) & 0&0&0&0&0&0&0&1&1&1&1&1&1&1&1
        \\
        (1,2) & 0&0&0&0&0&0&0&0&0&0&0&1&1&1&1
        \\
        (1,3) & 0&0&0&0&0&0&0&0&0&1&1&0&0&1&1
        \\
        (1,4) & 0&0&0&0&0&0&0&0&1&0&1&0&1&0&1
        \\
        (2,2) & 0&0&0&1&1&1&1&0&0&0&0&1&1&1&1
        \\
        (2,3) & 0&0&0&0&0&1&1&0&0&0&0&0&0&1&1
        \\
        (2,4) & 0&0&0&0&1&0&1&0&0&0&0&0&1&0&1
        \\
        (3,3) & 0&1&1&0&0&1&1&0&0&1&1&0&0&1&1
        \\
        (3,4) & 0&0&1&0&0&0&1&0&0&0&1&0&0&0&1
        \\
        (4,4) & 1&0&1&0&1&0&1&0&1&0&1&0&1&0&1
      \end{NiceArray}}
    \label{eq:illustration_build_Pk}
  \end{IEEEeqnarray*}
  Every row of $\mat{H}_{\tn{X}}$ has weight $8$, every pair of rows overlaps in $4$ positions, and every triple in $2$; hence $\mat{P}_4\vect{x}=2\trans{(4,2,2,2,4,2,2,4,2,4)}$ and $\mat{T}_4\vect{x}=2\trans{(1,1,1,1)}$, so $(\tn{O1})$ and $(\tn{O2})$ hold. All $15$ nonzero codewords have weight $8$, so $\delta_{i,8}=1$ $\forall i$, which satisfies $(\tn{W1})$ and $(\tn{W2})$. Since $\mathrm{K}_1(8;15)=-1$, $\mathrm{K}_2(8;15)=-7$, and $\mathrm{K}_3(8;15)=7$, the left-hand side of $(\tn{D=})$ equals $-15=-\binom{15}{1}$ for $\ell=1$ and $-105=-\binom{15}{2}$ for $\ell=2$, so $B_1=B_2=0$ and $B_3=35$. Thus $\dual{d}=3$, and appending the all-ones row gives the $\llbracket 15,1,3\rrbracket$ triorthogonal code.
\end{example}
}

\subsection{Search for Nontrivial Triorthogonal Codes}
\label{sec:search_nontrivial-TriOrthogonal-codes}

Solving the ILP with all vectors in $\set{V}$ is computationally costly. To reduce the size of the formulation, we impose a prescribed symmetry on the column types, following the ideas of~\cite[Th.~3]{KohnertWassermann09_1} and~\cite[Sec.~23.4]{HuffmanKimSole21_1}. Specifically, we prescribe a group of automorphisms $\Gamma$ acting on the binary column types in $\set{V}$. \ifthenelse{\boolean{arXiv}}{For instance, a permutation automorphism $\pi$ of the code generated by $\mat{H}_{\tn{X}}$ acts on $\set{V}$ through the invertible matrix $\mat{A}_{\pi}\in\Field_2^{k_0\times k_0}$ satisfying $\mat{A}_{\pi}\mat{H}_{\tn{X}}=\mat{H}_{\tn{X}}\mat{Q}_\pi$, where $\mat{Q}_\pi$ is the permutation matrix of $\pi$.}{} This group action partitions $\set{V}$ into orbits, and instead of assigning one variable to each vector in $\set{V}$, we assign one variable to each orbit under $\Gamma$. This orbit-based reduction preserves the prescribed symmetry while substantially reducing the number of variables in the ILP. \ifthenelse{\boolean{arXiv}}{Each orbit variable gives the common multiplicity of all column types in its orbit, so an orbit $\set{O}$ contributes $|\set{O}|$ times its variable to $(\tn{L1})$. For a $\Gamma$-invariant multiplicity vector, $\Gamma$ also permutes the nonzero codewords while preserving their weights; hence, we likewise assign the indicators $\delta_{i,j}$ to codeword orbits, each weighted by its orbit size in $(\tn{D=})$.}{Each orbit variable gives the common multiplicity of all column types in its orbit, and the indicators $\delta_{i,j}$ are likewise assigned to codeword orbits. Due to space constraints, we do not report the full formulation here.}

Based on the approach suggested in~\cite[Sec.~23.4.1]{HuffmanKimSole21_1}, we first restrict the orbit formulation to cyclic automorphism groups and vary $k_0$ from $6$ to $10$. This provides a computationally tractable search space for nontrivial even-weight triorthogonal matrices with $\dual{d}\geq 3$. To target the quantum minimum distance $d=5$ directly, we note that only the odd-weight codewords of $\dual{\code{C}}$ correspond to logical $\mat{Z}$ operators, whereas its even-weight codewords are $\mat{Z}$-type stabilizers. Accordingly, we impose $(\tn{D=})$ only for $\ell\in\{1,3\}$, which guarantees $d_{\tn{X}}\geq 5$ while admitting $\mat{X}$-degenerate codes, and we restrict the weight indicators to $\delta_{i,j}$ with $j\leq n-5$, which guarantees $d_{\tn{Z}}\geq 5$. Extending the cyclic search to $k_0\in\{11,15,16\}$ yields three new codes, with parameters $\llbracket 89,1,5\rrbracket$, $\llbracket 93,1,5\rrbracket$, and $\llbracket 85,1,5\rrbracket$, respectively, and the quantum minimum distance of each feasible solution is verified exactly. The nontrivial triorthogonal codes obtained for odd lengths $n\leq 93$ are listed in Table~\ref{tab:code-parameters_nontrivial-TriOrthogonal-codes}, where codes that are also triply even are marked by ``TE''. Within this cyclic-symmetry search, all feasible solutions found have classical dual distance $\dual{d}\in\{3,4,5\}$; in particular, $d_0=4$ for the $\mat{X}$-degenerate $\llbracket 89,1,5\rrbracket$ code and $d_0=5$ for the other two codes with $d=5$. This indicates that cyclic symmetry alone may be too restrictive for obtaining higher-distance instances.

This motivates the use of larger automorphism groups. To validate this approach, we choose $\Gamma$ as the group generated by permutation automorphisms of  known codes. For the $\llbracket 49,1,5\rrbracket$ code of~\cite{BravyiHaah12_1}, the \ac{ILP} with $\dual{d}=4$ returns exactly this code. For the $\llbracket 95,1,7\rrbracket$ code of~\cite{Sullivan24_1, JainAlbert25_1}, obtained through the doubling construction of~\cite{BravyiCross15_1sub} and also $\mat{X}$-degenerate with $d_0=4$, the automorphisms described in Section~\ref{subsec:disc} split the $2^{25}-1$ column types into $89$ orbits; imposing $(\tn{D=})$ for $\ell\in\{1,2,3,5\}$, which guarantees $d_{\tn{X}}\geq 7$,  the \ac{ILP} recovers the original construction with $\dual{d}=4$.

\begin{table}[t!]
\centering
\renewcommand{\arraystretch}{1.0}
\caption{Triorthogonal Codes Found by the Proposed ILP Formulation}
 \vspace{-2ex}
\begin{tabular}{llll}
\toprule
\multicolumn{4}{c}{\textbf{Triorthogonal codes}}  \\
\midrule
$\llbracket 39, 1, 3 \rrbracket$ & $\llbracket 43, 1, 3 \rrbracket$ &
$\llbracket 45, 1, 3 \rrbracket$ (\tn{TE}) & $\llbracket 47, 1, 3 \rrbracket$ (\tn{TE}) \\
$\llbracket 49, 1, 3 \rrbracket$ & $\llbracket 51, 1, 3 \rrbracket$ &
$\llbracket 53, 1, 3 \rrbracket$ & $\llbracket 55, 1, 3 \rrbracket$ \\
$\llbracket 57, 1, 3 \rrbracket$ & $\llbracket 59, 1, 3 \rrbracket$ &
$\llbracket 61, 1, 3 \rrbracket$ & $\llbracket 63, 1, 3 \rrbracket$ (\tn{TE}) \\
$\llbracket 65, 1, 3 \rrbracket$ & $\llbracket 67, 1, 3 \rrbracket$ &
$\llbracket 69, 1, 3 \rrbracket$ (\tn{TE}) & $\llbracket 71, 1, 3 \rrbracket$ (\tn{TE}) \\
$\llbracket 73, 1, 3 \rrbracket$ & $\llbracket 75, 1, 3 \rrbracket$ &
$\llbracket 77, 1, 3 \rrbracket$ & $\llbracket 79, 1, 3 \rrbracket$ \\
$\llbracket 85, 1, 5 \rrbracket$ (\tn{TE}) & $\llbracket 89, 1, 5 \rrbracket$ (\tn{TE})\textsuperscript{a} &
$\llbracket 93, 1, 5 \rrbracket$ (\tn{TE}) & -- \\
\bottomrule
\end{tabular}
\par\smallskip
{\raggedright\footnotesize \textsuperscript{a}\,$\mat{X}$-degenerate, with $d_0=4<d_{\tn{X}}=5$.\par}
\label{tab:code-parameters_nontrivial-TriOrthogonal-codes}
\end{table}

\section{Numerical Results}
\label{sec:num_res}

\ifthenelse{\boolean{arXiv}}{%
\begin{figure*}[t]
    \centering
    \subfloat[]{
        \resizebox{\TriPlotWidth}{!}{
            \begin{tikzpicture} 
    \begin{loglogaxis}[
        xlabel={$p$},
        ylabel={LER},
        xmin=0.001, xmax=0.1,
        ymin=0.00005, ymax=1,
        grid=both,
        major grid style={solid, gray!70},
        minor grid style={solid, gray!30},
        legend cell align=left,
        legend style={legend columns=1, font=\tiny, inner sep=1pt, row sep=-1pt, fill opacity=0.85, text opacity=1},
        legend pos = south east,
        height  = 7.5cm,
        width = \columnwidth, 
        scaled x ticks=false,
        tri axis=leg89
    ]

    \addplot[
        color = ForestGreen, 
        mark = o, densely dashed, thick,
        mark size = 2.5pt,
        mark options = {solid}
        ]
        coordinates {
            (0.001, 1.9216e-02)
            (0.002, 4.1760e-02)
            (0.003, 6.6445e-02)
            (0.004, 9.4215e-02)
            (0.005, 1.2389e-01)
            (0.006, 1.5383e-01)
            (0.007, 1.8640e-01)
            (0.008, 2.1834e-01)
            (0.009, 2.4904e-01)
            (0.01, 2.8141e-01)
            (0.02, 5.6193e-01)
            (0.03, 7.5035e-01)
            (0.04, 8.5301e-01)
            (0.05, 9.0770e-01)
            (0.06, 9.3409e-01)
            (0.07, 9.4608e-01)
            (0.08, 9.5220e-01)
            (0.09, 9.5447e-01)
            (0.1, 9.5545e-01)
    };

    \addplot[ 
        color = ForestGreen, 
        mark = *, densely dashed, thick,
        mark size = 2.5pt,
        mark options = {solid}
        ]
        coordinates {
            (0.1, 0.49504950495049505)
            (0.09, 0.4807692307692308)
            (0.08, 0.45871559633027525)
            (0.07, 0.5102040816326531)
            (0.06, 0.49261083743842365)
            (0.05, 0.45871559633027525)
            (0.04, 0.352112676056338)
            (0.03, 0.30303030303030304)
            (0.02, 0.1937984496124031)
            (0.01, 0.07087172218284904)
            (0.009, 0.060350030175015085)
            (0.008, 0.042662116040955635)
            (0.007, 0.036231884057971016)
            (0.006, 0.023084025854108958)
            (0.005, 0.01688048615800135)
            (0.004, 0.012132977432661975)
            (0.003, 0.006401638819537802)
            (0.002, 0.0031007751937984496)
            (0.001, 0.0006913382233990335)
            (0.0009, 0.0005859294892452642)
            (0.0008, 0.00043756782301256693)
            (0.0007, 0.0003739240335933352)
            (0.0006, 0.00022585242350943048)
            (0.0005, 0.00020497809809021905)
            (0.0004, 0.0001057713055140697)
            (0.0003, 6.374993704693717e-05)
            (0.0002, 2.9788022474467196e-05)
            (0.0001, 7.065727160478562e-06)
    };

    \addplot[ 
        color = ForestGreen, 
        mark = *, thick,
        mark size = 2.5pt,
        mark options = {solid}
        ]
        coordinates {
            (0.1, 0.5434782608695652)
            (0.09, 0.4444444444444444)
            (0.08, 0.4784688995215311)
            (0.07, 0.4672897196261682)
            (0.06, 0.44052863436123346)
            (0.05, 0.46511627906976744)
            (0.04, 0.398406374501992)
            (0.03, 0.303951367781155)
            (0.02, 0.18214936247723132)
            (0.01, 0.056274620146314014)
            (0.009, 0.037993920972644375)
            (0.008, 0.03636363636363636)
            (0.007, 0.02795638803466592)
            (0.006, 0.019550342130987292)
            (0.005, 0.012965123816932453)
            (0.004, 0.00905961224859576)
            (0.003, 0.003512839428109741)
            (0.002, 0.0013108049653292088)
            (0.001, 0.000339914069723174)
            (0.0009, 0.00029916204710605593)
            (0.0008, 0.0002074529548561625)
            (0.0007, 0.00014776745540008777)
            (0.0006, 9.046695423095852e-05)
            (0.0005, 7.646452504824912e-05)
            (0.0004, 3.9397439796772245e-05)
            (0.0003, 2.329354950368434e-05)
            (0.0002, 1.1147508966498836e-05)
            (0.0001, 2.701873884632687e-06)
        };

    \addplot[
        color = RedOrange, %
        mark = triangle*, densely dashed, thick,
        mark size = 2.5pt,
        mark options = {solid}
        ] 
        coordinates {
            (0.001, 2.1500e-03)
            (0.002, 9.2100e-03)
            (0.003, 1.8975e-02)
            (0.004, 3.2945e-02)
            (0.005, 4.8275e-02)
            (0.006, 6.6175e-02)
            (0.007, 8.7235e-02)
            (0.008, 1.0946e-01)
            (0.009, 1.3228e-01)
            (0.010, 1.5422e-01)
            (0.020, 3.9177e-01)
            (0.030, 5.7664e-01)
            (0.040, 6.8658e-01)
            (0.050, 7.4861e-01)
            (0.060, 7.8073e-01)
            (0.070, 7.9609e-01)
            (0.080, 8.0562e-01)
            (0.090, 8.0675e-01)
            (0.100, 8.0838e-01)
        };

    \addplot[
        color = RedOrange, %
        mark = triangle*, thick, %
        mark size = 2.5pt,
        mark options = {solid}
        ] 
        coordinates {
            (0.001, 9.0000e-05)
            (0.002, 7.7500e-04)
            (0.003, 2.2900e-03)
            (0.004, 5.0800e-03)
            (0.005, 9.0550e-03)
            (0.006, 1.3895e-02)
            (0.007, 2.0930e-02)
            (0.008, 2.8785e-02)
            (0.009, 3.7895e-02)
            (0.010, 4.8295e-02)
            (0.020, 1.8156e-01)
            (0.030, 3.1105e-01)
            (0.040, 3.9749e-01)
            (0.050, 4.4906e-01)
            (0.060, 4.7436e-01)
            (0.070, 4.8881e-01)
            (0.080, 4.9752e-01)
            (0.090, 4.9704e-01)
            (0.100, 4.9896e-01)
        };

    \addplot[ 
        color = blue, 
        mark = asterisk, thick, densely dashed,
        mark size = 2.5pt,
        mark options = {solid}
        ]
        coordinates {
            (0.100000, 5.848e-01)
            (0.090000, 4.762e-01)
            (0.080000, 5.263e-01)
            (0.070000, 4.425e-01)
            (0.060000, 5.376e-01)
            (0.050000, 3.937e-01)
            (0.040000, 4.098e-01)
            (0.030000, 2.825e-01)
            (0.020000, 2.016e-01)
            (0.010000, 4.762e-02)
            (0.009000, 4.016e-02)
            (0.008000, 2.881e-02)
            (0.007000, 2.230e-02)
            (0.006000, 1.326e-02)
            (0.005000, 9.376e-03)
            (0.004000, 4.946e-03)
            (0.003000, 2.092e-03)
            (0.002000, 6.983e-04)
            (0.001000, 1.233e-04)
            (0.000900, 6.651e-05)
            (0.000800, 5.316e-05)
            (0.000700, 4.297e-05)
            (0.000600, 2.378e-05)
            (0.000500, 1.331e-05)
            (0.000400, 6.544e-06)
            (0.000300, 2.373e-06)
            (0.000200, 8.225e-07)
            (0.000100, 1.053e-07)
    };

    \legend{
        {\texttt{BP}$2$}, %
        {\texttt{BP}$2+$\texttt{OSD-}$10$},
        {\texttt{BP}$2+$\texttt{OSD-}$60$},
        {\texttt{AED-}$8$}, %
        {\texttt{AED-}$8+$\texttt{OSD-}$1$}, %
        {\texttt{qGRAND-}$10^{7}$},
    }

     \end{loglogaxis}
\end{tikzpicture}
        }
        \label{fig:89_LER}
    }
    \subfloat[]{
        \resizebox{\TriPlotWidth}{!}{
            \definecolor{curvegray}{RGB}{125,125,125}
\definecolor{AEDorange}{RGB}{235,85,25}
\definecolor{AEDpurple}{RGB}{145,70,180}
\definecolor{AEDteal}{RGB}{0,155,150}
\definecolor{qblue}{RGB}{0,75,215}
\definecolor{newAEDgreen}{RGB}{0,135,85}
\definecolor{newAEDblue}{RGB}{20,110,175}
\definecolor{newAEDred}{RGB}{190,55,55}

\begin{tikzpicture}
\begin{loglogaxis}[
    xlabel={$p$},
    ylabel={ $\eta$
    },
    xmin=1e-3,
    xmax=1e-1,
    ymin=1e4,
    ymax=5e7,
    grid=both,
    major grid style={solid,gray!70},
    minor grid style={solid,gray!30},
    height=7.5cm,
    width=\columnwidth,
    legend cell align=left,
    legend pos = south east,
    legend style={legend columns=1, font=\tiny, inner sep=1pt, row sep=-1pt, fill opacity=0.85, text opacity=1},
    tri cost axis,
]

\addplot[
    color= RedOrange, %
    very thick
] coordinates {
    (0.001,4.87133e+06)
    (0.100,4.87133e+06)
};
\aedentry{\texttt{AED-}$8+$\texttt{OSD-}$1$ (full evaluation)}{full}

\addplot[cost noleg, color=ForestGreen,mark=*,mark size = 2.5pt,thick, mark options = {solid}] coordinates {
    (0.001,723707) (0.002,991346) (0.003,1236229)
    (0.004,1464190) (0.005,1655297) (0.006,1845561)
    (0.007,2056456) (0.008,2182088) (0.009,2316526)
    (0.010,2426748) (0.020,3196682) (0.030,3609186)
    (0.040,3736495) (0.050,3772228) (0.060,3803533)
    (0.070,3803533) (0.080,3803533) (0.090,3803533)
    (0.100,3803533)
};
\costentry{\texttt{BP}$2+$\texttt{OSD-}$60$}

\addplot[
    color= RedOrange, %
    mark=triangle*, densely dotted,
    mark size=2.5pt,    
    thick,
    mark options={solid}
] coordinates {
 (0.001,701580)
    (0.002,943597)
    (0.003,1.16625e+06)
    (0.004,1.37109e+06)
    (0.005,1.55956e+06)
    (0.006,1.73296e+06)
    (0.007,1.89251e+06)
    (0.008,2.03932e+06)
    (0.009,2.1744e+06)
    (0.01,2.2987e+06)
    (0.02,3.10775e+06)
    (0.03,3.45903e+06)
    (0.04,3.6106e+06)
    (0.05,3.67533e+06)
    (0.06,3.7026e+06)
    (0.07,3.7139e+06)
    (0.08,3.7185e+06)
    (0.09,3.72033e+06)
    (0.1,3.72104e+06)
};
\aedentry{\texttt{AED-}$8+$\texttt{OSD-}$1$ (progressive)}{progressive}

\addplot[cost noleg, color=ForestGreen,mark=*,densely dashed,mark size = 2.5pt,thick, mark options = {solid}] coordinates {
    (0.001,459613) (0.002,479626) (0.003,497937)
    (0.004,514983) (0.005,529273) (0.006,543500)
    (0.007,559270) (0.008,568664) (0.009,578716)
    (0.010,586958) (0.020,644530) (0.030,675375)
    (0.040,684895) (0.050,687567) (0.060,689908)
    (0.070,689908) (0.080,689908) (0.090,689908)
    (0.100,689908)
};
\costentry{\texttt{BP}$2+$\texttt{OSD-}$10$}

\addplot[ 
        cost noleg,
        color = blue, 
        mark = asterisk, thick, densely dashed,
        mark size = 2.5pt,
        mark options = {solid}
        ]
        coordinates {
    (0.001,18718) (0.002,51632) (0.003,97547)
    (0.004,156385) (0.005,223094) (0.006,303862)
    (0.007,415800) (0.008,457839) (0.009,565274)
    (0.010,688315) (0.020,1650789) (0.030,2021028)
    (0.040,2298241) (0.050,2672395) (0.060,3173565)
    (0.070,2829133) (0.080,2844171) (0.090,2846853)
    (0.100,3058689)
};
\costentry{\texttt{qGRAND-}$10^{7}$}

\end{loglogaxis}

\end{tikzpicture}
        }
        \label{fig:costs_89}
    }
    \caption{(a) \acs{LER} of the $\llbracket 89, 1, 5 \rrbracket$ triorthogonal code on the dephasing channel using several decoders. 
    (b) Comparison of the estimated average number of binary operations $\eta$ per decoded frame for the $\llbracket 89, 1, 5 \rrbracket$ triorthogonal code considering several decoders.} 
    \label{fig:Tri_89}
\end{figure*}

\begin{figure*}[t]
    \centering
    \subfloat[]{
        \resizebox{\TriPlotWidth}{!}{
            \begin{tikzpicture} 
    \begin{loglogaxis}[
        xlabel={$p$},
        ylabel={LER},
        xmin=0.001, xmax=0.1,
        ymin=0.0000005, ymax=1,
        grid=both,
        major grid style={solid, gray!70},
        minor grid style={solid, gray!30},
        legend pos = south east,
        height  = 7.5cm,
        width = \columnwidth, 
        legend cell align=left,
        legend style={legend columns=1, font=\tiny, inner sep=1pt, row sep=-1pt, fill opacity=0.85, text opacity=1},
        scaled x ticks=false,
        tri axis=leg95
    ]

    \addplot[ 
        color = ForestGreen, 
        mark = o, thick, densely dashed,
        mark size = 2.5pt,
        mark options = {solid},
        ]
        coordinates {
            (0.001, 2.2954e-02)
            (0.002, 4.5851e-02)
            (0.003, 6.8724e-02)
            (0.004, 9.1593e-02)
            (0.005, 1.1446e-01)
            (0.006, 1.3732e-01)
            (0.007, 1.6016e-01)
            (0.008, 1.8296e-01)
            (0.009, 2.0568e-01)
            (0.01, 2.2831e-01)
            (0.02, 4.4232e-01)
            (0.03, 6.1792e-01)
            (0.04, 7.4707e-01)
            (0.05, 8.3536e-01)
            (0.06, 8.9299e-01)
            (0.07, 9.2961e-01)
            (0.08, 9.5259e-01)
            (0.09, 9.6696e-01)
            (0.1, 9.7595e-01)
    };
    \addlegendentry{\texttt{BP}$2$} %

    \addplot[ 
        color = ForestGreen, 
        mark = *, thick, densely dashed,
        mark size = 2.5pt,
        mark options = {solid},
        ]
        coordinates {
            (0.1, 0.4694835680751174)
            (0.09, 0.5154639175257731)
            (0.08, 0.47619047619047616)
            (0.07, 0.4219409282700422)
            (0.06, 0.4166666666666667)
            (0.05, 0.33112582781456956)
            (0.04, 0.24752475247524752)
            (0.03, 0.14792899408284024)
            (0.02, 0.08992805755395683)
            (0.01, 0.021767522855899)
            (0.009, 0.01881821603312006)
            (0.008, 0.015554518587649712)
            (0.007, 0.01038961038961039)
            (0.006, 0.00804246421103426)
            (0.005, 0.005420347986340723)
            (0.004, 0.0037054878274724867)
            (0.003, 0.0019806267806267807)
            (0.002, 0.000985017878074487)
            (0.001, 0.0002574161595568323)
    };
    \addlegendentry{\texttt{BP}$2+$\texttt{OSD-}$10$}

    \addplot[ 
        color = ForestGreen, 
        mark = *, thick,
        mark size = 2.5pt,
        mark options = {solid},
        ]
        coordinates {
            (0.09, 0.47619047619047616)
            (0.08, 0.4484304932735426)
            (0.07, 0.3508771929824561)
            (0.06, 0.4149377593360996)
            (0.05, 0.27472527472527475)
            (0.04, 0.17064846416382254)
            (0.03, 0.11363636363636363)
            (0.02, 0.03544842254519674)
            (0.01, 0.004672897196261682)
            (0.009, 0.002952378140592247)
            (0.008, 0.002541231480775584)
            (0.007, 0.0013012361743656475)
            (0.006, 0.000877108349194376)
            (0.005, 0.00042288483577267405)
            (0.004, 0.00021216602415722352)
            (0.003, 8.471668622210173e-05)
            (0.002, 2.4135741340156345e-05)
            (0.001, 2.6332934292824703e-06)
    };
     \addlegendentry{\texttt{BP}$2+$\texttt{OSD-}$60$}

    \addplot[ 
        color = RedOrange, 
        mark = triangle*, thick, densely dashed,
        mark size = 2.5pt,
        mark options = {solid},
        ]
        coordinates {
            (0.001, 5.9507e-03)
            (0.002, 1.1892e-02)
            (0.003, 1.7944e-02)
            (0.004, 2.4210e-02)
            (0.005, 3.0780e-02)
            (0.006, 3.7725e-02)
            (0.007, 4.5105e-02)
            (0.008, 5.2964e-02)
            (0.009, 6.1335e-02)
            (0.01, 7.0240e-02)
            (0.02, 1.8757e-01)
            (0.03, 3.3686e-01)
            (0.04, 4.8526e-01)
            (0.05, 6.1221e-01)
            (0.06, 7.1133e-01)
            (0.07, 7.8461e-01)
            (0.08, 8.3718e-01)
            (0.09, 8.7438e-01)
            (0.1, 9.0053e-01)
    };
    \addlegendentry{$\texttt{AED-}8$} %

    \addplot[ 
        color = RedOrange, 
        mark = triangle*, thick,
        mark size = 2.5pt,
        mark options = {solid},
        ]
        coordinates {
            (0.001, 4.9476e-07)
            (0.002, 7.5043e-06)
            (0.003, 3.6025e-05)
            (0.004, 1.0800e-04)
            (0.005, 2.5017e-04)
            (0.006, 4.9231e-04)
            (0.007, 8.6579e-04)
            (0.008, 1.4024e-03)
            (0.009, 2.1332e-03)
            (0.01, 3.0883e-03)
            (0.02, 2.9749e-02)
            (0.03, 9.1422e-02)
            (0.04, 1.7644e-01)
            (0.05, 2.6508e-01)
            (0.06, 3.4209e-01)
            (0.07, 4.0070e-01)
            (0.08, 4.4094e-01)
            (0.09, 4.6641e-01)
            (0.1, 4.8160e-01)
    };
    \addlegendentry{$\texttt{AED-}8+$\texttt{OSD-}$1$} %
    
    \addplot[ 
        color = blue, 
        mark = asterisk, thick,
        mark size = 2.5pt,
        mark options = {solid}
        ]
        coordinates {
            (0.0005, 1.2546e-07)
            (0.0006, 2.1192e-07)
            (0.0007, 4.2244e-07)
            (0.0008, 6.3807e-07)
            (0.0009, 1.2689e-06)
            (0.0010, 1.4929e-06)
            (0.0015, 8.4011e-06)
            (0.0020, 2.613e-05)
            (0.0030, 0.00014198)
            (0.0040, 0.00041927)
            (0.0050, 0.00079025)
            (0.0060, 0.0016046)
            (0.0070, 0.0032247)
            (0.0080, 0.0042012)
            (0.0090, 0.0062858)
            (0.0100, 0.0099711)
            (0.0150, 0.038432)
            (0.0200, 0.084459)
            (0.0250, 0.14556)
            (0.0300, 0.22676)
            (0.0400, 0.4386)
            (0.0500, 0.54645)
            (0.0600, 0.66667)
            (0.0700, 0.72464)
            (0.0800, 0.84746)
            (0.0900, 0.90909)
            (0.1000, 0.95238)
    };
    \addlegendentry{$\texttt{qGRAND-}10^{6}$}

    \addplot[ 
        color = blue, 
        mark = asterisk, thick, dashed,
        mark size = 2.5pt,
        mark options = {solid},
        ]
        coordinates {
            (0.0006, 5.200e-08) %
            (0.0007, 1.064e-07)
            (0.0008, 1.704e-07)
            (0.0009, 3.039e-07)
            (0.0010, 5.035e-07)
            (0.0015, 2.2263e-06)
            (0.0020, 7.454e-06)
            (0.0030, 4.1446e-05)
            (0.0040, 0.00010927)
            (0.0050, 0.00028793)
            (0.0060, 0.00056864)
            (0.0070, 0.0010685)
            (0.0080, 0.0020293)
            (0.0090, 0.0026455)
            (0.0100, 0.0037703)
            (0.0150, 0.016263)
            (0.0200, 0.038971)
            (0.0250, 0.08285)
            (0.0300, 0.13193)
            (0.0400, 0.27778)
            (0.0500, 0.49751)
            (0.0600, 0.5291)
            (0.0700, 0.71429)
            (0.0800, 0.72993)
            (0.0900, 0.81967)
            (0.1000, 0.91743)
    };
    \addlegendentry{$\texttt{qGRAND-}10^{7}$}

    \end{loglogaxis}
\end{tikzpicture}
        }
        \label{fig:95_LER}
    }
    \subfloat[]{
        \resizebox{\TriPlotWidth}{!}{
            \definecolor{newAEDgreen}{RGB}{0,135,85}
\definecolor{newAEDpurple}{RGB}{145,70,180}

\begin{tikzpicture}
    \begin{loglogaxis}[
        xlabel={$p$},
        ylabel={$\eta$},
        xmin=0.001,
        xmax=0.1,
        ymin=4e3,
        ymax=1e11,
        grid=both,
        major grid style={solid, gray!70},
        minor grid style={solid, gray!30},
        legend pos = south east,
        height  = 7.5cm,
        width = \columnwidth,
        legend cell align=left,
        legend style = {legend columns=1, font=\tiny, inner sep=1pt, row sep=-1pt, fill opacity=0.85, text opacity=1}, %
        scaled x ticks=false,
        tri cost axis,
        ytick={1e4,1e6,1e8,1e10}
    ]

    \addplot[
        color= RedOrange, %
        very thick
    ] coordinates {
        (0.001,6216136.7221)
        (0.100,6216136.7221)
    };
    \aedentry{\texttt{AED-}$8+$\texttt{OSD-}$1$ (full evaluation)}{full}

    \addplot[
        color=RedOrange, %
        mark=triangle*, densely dotted,
        thick,
        mark size=2.5pt,
        mark options={solid}
    ] coordinates {
        (0.001,633296.695146) (0.002,860439.959722)
        (0.003,1067851.349140) (0.004,1257612.435640)
        (0.005,1431681.409290) (0.006,1591881.255660)
        (0.007,1739893.218680) (0.008,1877254.461760)
        (0.009,2005359.015440) (0.010,2125461.250780)
        (0.020,3076075.245650) (0.030,3852342.426350)
        (0.040,4530063.491980) (0.050,5073900.853880)
        (0.060,5472389.927320) (0.070,5744990.003230)
        (0.080,5922920.087400) (0.090,6035560.964970)
        (0.100,6105348.154670)
    };
    \aedentry{\texttt{AED-}$8+$\texttt{OSD-}$1$ (progressive)}{progressive}

    \addplot[
        cost noleg,
        color=ForestGreen,
        mark=*,
        thick,
        mark size=2.5pt,
        mark options={solid}
    ]
    coordinates {
        (0.1, 7138941)
        (0.09, 7138941)
        (0.08, 7138941)
        (0.07, 7115235)
        (0.06, 7138941)
        (0.05, 7120380)
        (0.04, 7035180)
        (0.03, 6755081)
        (0.02, 6130700)
        (0.01, 4562847)
        (0.009, 4268104)
        (0.008, 3985452)
        (0.007, 3678251)
        (0.006, 3330326)
        (0.005, 2941093)
        (0.004, 2525022)
        (0.003, 2065289)
        (0.002, 1552158)
        (0.001, 995829)
    };
    \costentry{\texttt{BP}$2+$\texttt{OSD-}$60$}

    \addplot[
        cost noleg,
        color=blue,
        mark=asterisk,
        thick,
        mark size = 2.5pt,
        mark options={solid}
    ]
    coordinates {
        (0.0007, 57000)
        (0.0008, 71250)
        (0.0009, 90250)
        (0.001,  109250)
        (0.002,  584250)
        (0.003,  1890500)
        (0.004,  4365250)
        (0.005,  7742500)
        (0.006,  13775000)
        (0.007,  23289250)
        (0.008,  29972500)
        (0.009,  45429000)
        (0.010,  68666000)
        (0.020,  465186500)
        (0.030,  1151922500)
        (0.040,  2060502500)
        (0.050,  2584427500)
        (0.060,  3185635000)
        (0.070,  3192522500)
        (0.080,  4000450000)
        (0.090,  4377932500)
        (0.100,  4310245000)
    };
    \costentry{\texttt{qGRAND-}$10^{6}$}

    \addplot[
        cost noleg,
        color=blue,
        mark=asterisk,
        thick,
        densely dashed,
        mark size = 2.5pt,
        mark options={solid}
    ]
    coordinates {
        (0.0007, 57000)
        (0.0008, 76000)
        (0.0009, 95000)
        (0.001,  118750)
        (0.002,  693500)
        (0.003,  2569750)
        (0.004,  6683250)
        (0.005,  13646750)
        (0.006,  25056250)
        (0.007,  47105750)
        (0.008,  69055500)
        (0.009,  107625500)
        (0.010,  151710250)
        (0.020,  1505797500)
        (0.030,  4743160000)
        (0.040,  11126875000)
        (0.050,  17939325000)
        (0.060,  22757250000)
        (0.070,  32246325000)
        (0.080,  33545450000)
        (0.090,  36166500000)
        (0.100,  39991675000)
    };
    \costentry{\texttt{qGRAND-}$10^{7}$}

    \end{loglogaxis}
\end{tikzpicture}
        }
        \label{fig:costs_95}
    }
    \caption{(a) \acs{LER} of the $\llbracket 95, 1, 7 \rrbracket$ triorthogonal code on the dephasing channel using several decoders. 
    (b) Comparison of the estimated average number of binary operations $\eta$ per decoded frame for the $\llbracket 95, 1, 7 \rrbracket$ triorthogonal code considering several decoders.
    }
    \label{fig:Tri95}
\end{figure*}
}{%
\begin{figure*}[t]
  \centering
  \begin{minipage}[t]{0.97\textwidth}\centering
    \subfloat[]{\makebox[0.245\linewidth][r]{\begin{tikzpicture} 
    \begin{loglogaxis}[
        xlabel={$p$},
        ylabel={LER},
        xmin=0.001, xmax=0.1,
        ymin=0.00005, ymax=1,
        grid=both,
        major grid style={solid, gray!70},
        minor grid style={solid, gray!30},
        legend cell align=left,
        legend style={legend columns=1, font=\tiny, inner sep=1pt, row sep=-1pt, fill opacity=0.85, text opacity=1},
        legend pos = south east,
        height  = 7.5cm,
        width = \columnwidth, 
        scaled x ticks=false,
        tri axis=leg89
    ]

    \addplot[
        color = ForestGreen, 
        mark = o, densely dashed, thick,
        mark size = 2.5pt,
        mark options = {solid}
        ]
        coordinates {
            (0.001, 1.9216e-02)
            (0.002, 4.1760e-02)
            (0.003, 6.6445e-02)
            (0.004, 9.4215e-02)
            (0.005, 1.2389e-01)
            (0.006, 1.5383e-01)
            (0.007, 1.8640e-01)
            (0.008, 2.1834e-01)
            (0.009, 2.4904e-01)
            (0.01, 2.8141e-01)
            (0.02, 5.6193e-01)
            (0.03, 7.5035e-01)
            (0.04, 8.5301e-01)
            (0.05, 9.0770e-01)
            (0.06, 9.3409e-01)
            (0.07, 9.4608e-01)
            (0.08, 9.5220e-01)
            (0.09, 9.5447e-01)
            (0.1, 9.5545e-01)
    };

    \addplot[ 
        color = ForestGreen, 
        mark = *, densely dashed, thick,
        mark size = 2.5pt,
        mark options = {solid}
        ]
        coordinates {
            (0.1, 0.49504950495049505)
            (0.09, 0.4807692307692308)
            (0.08, 0.45871559633027525)
            (0.07, 0.5102040816326531)
            (0.06, 0.49261083743842365)
            (0.05, 0.45871559633027525)
            (0.04, 0.352112676056338)
            (0.03, 0.30303030303030304)
            (0.02, 0.1937984496124031)
            (0.01, 0.07087172218284904)
            (0.009, 0.060350030175015085)
            (0.008, 0.042662116040955635)
            (0.007, 0.036231884057971016)
            (0.006, 0.023084025854108958)
            (0.005, 0.01688048615800135)
            (0.004, 0.012132977432661975)
            (0.003, 0.006401638819537802)
            (0.002, 0.0031007751937984496)
            (0.001, 0.0006913382233990335)
            (0.0009, 0.0005859294892452642)
            (0.0008, 0.00043756782301256693)
            (0.0007, 0.0003739240335933352)
            (0.0006, 0.00022585242350943048)
            (0.0005, 0.00020497809809021905)
            (0.0004, 0.0001057713055140697)
            (0.0003, 6.374993704693717e-05)
            (0.0002, 2.9788022474467196e-05)
            (0.0001, 7.065727160478562e-06)
    };

    \addplot[ 
        color = ForestGreen, 
        mark = *, thick,
        mark size = 2.5pt,
        mark options = {solid}
        ]
        coordinates {
            (0.1, 0.5434782608695652)
            (0.09, 0.4444444444444444)
            (0.08, 0.4784688995215311)
            (0.07, 0.4672897196261682)
            (0.06, 0.44052863436123346)
            (0.05, 0.46511627906976744)
            (0.04, 0.398406374501992)
            (0.03, 0.303951367781155)
            (0.02, 0.18214936247723132)
            (0.01, 0.056274620146314014)
            (0.009, 0.037993920972644375)
            (0.008, 0.03636363636363636)
            (0.007, 0.02795638803466592)
            (0.006, 0.019550342130987292)
            (0.005, 0.012965123816932453)
            (0.004, 0.00905961224859576)
            (0.003, 0.003512839428109741)
            (0.002, 0.0013108049653292088)
            (0.001, 0.000339914069723174)
            (0.0009, 0.00029916204710605593)
            (0.0008, 0.0002074529548561625)
            (0.0007, 0.00014776745540008777)
            (0.0006, 9.046695423095852e-05)
            (0.0005, 7.646452504824912e-05)
            (0.0004, 3.9397439796772245e-05)
            (0.0003, 2.329354950368434e-05)
            (0.0002, 1.1147508966498836e-05)
            (0.0001, 2.701873884632687e-06)
        };

    \addplot[
        color = RedOrange, %
        mark = triangle*, densely dashed, thick,
        mark size = 2.5pt,
        mark options = {solid}
        ] 
        coordinates {
            (0.001, 2.1500e-03)
            (0.002, 9.2100e-03)
            (0.003, 1.8975e-02)
            (0.004, 3.2945e-02)
            (0.005, 4.8275e-02)
            (0.006, 6.6175e-02)
            (0.007, 8.7235e-02)
            (0.008, 1.0946e-01)
            (0.009, 1.3228e-01)
            (0.010, 1.5422e-01)
            (0.020, 3.9177e-01)
            (0.030, 5.7664e-01)
            (0.040, 6.8658e-01)
            (0.050, 7.4861e-01)
            (0.060, 7.8073e-01)
            (0.070, 7.9609e-01)
            (0.080, 8.0562e-01)
            (0.090, 8.0675e-01)
            (0.100, 8.0838e-01)
        };

    \addplot[
        color = RedOrange, %
        mark = triangle*, thick, %
        mark size = 2.5pt,
        mark options = {solid}
        ] 
        coordinates {
            (0.001, 9.0000e-05)
            (0.002, 7.7500e-04)
            (0.003, 2.2900e-03)
            (0.004, 5.0800e-03)
            (0.005, 9.0550e-03)
            (0.006, 1.3895e-02)
            (0.007, 2.0930e-02)
            (0.008, 2.8785e-02)
            (0.009, 3.7895e-02)
            (0.010, 4.8295e-02)
            (0.020, 1.8156e-01)
            (0.030, 3.1105e-01)
            (0.040, 3.9749e-01)
            (0.050, 4.4906e-01)
            (0.060, 4.7436e-01)
            (0.070, 4.8881e-01)
            (0.080, 4.9752e-01)
            (0.090, 4.9704e-01)
            (0.100, 4.9896e-01)
        };

    \addplot[ 
        color = blue, 
        mark = asterisk, thick, densely dashed,
        mark size = 2.5pt,
        mark options = {solid}
        ]
        coordinates {
            (0.100000, 5.848e-01)
            (0.090000, 4.762e-01)
            (0.080000, 5.263e-01)
            (0.070000, 4.425e-01)
            (0.060000, 5.376e-01)
            (0.050000, 3.937e-01)
            (0.040000, 4.098e-01)
            (0.030000, 2.825e-01)
            (0.020000, 2.016e-01)
            (0.010000, 4.762e-02)
            (0.009000, 4.016e-02)
            (0.008000, 2.881e-02)
            (0.007000, 2.230e-02)
            (0.006000, 1.326e-02)
            (0.005000, 9.376e-03)
            (0.004000, 4.946e-03)
            (0.003000, 2.092e-03)
            (0.002000, 6.983e-04)
            (0.001000, 1.233e-04)
            (0.000900, 6.651e-05)
            (0.000800, 5.316e-05)
            (0.000700, 4.297e-05)
            (0.000600, 2.378e-05)
            (0.000500, 1.331e-05)
            (0.000400, 6.544e-06)
            (0.000300, 2.373e-06)
            (0.000200, 8.225e-07)
            (0.000100, 1.053e-07)
    };

    \legend{
        {\texttt{BP}$2$}, %
        {\texttt{BP}$2+$\texttt{OSD-}$10$},
        {\texttt{BP}$2+$\texttt{OSD-}$60$},
        {\texttt{AED-}$8$}, %
        {\texttt{AED-}$8+$\texttt{OSD-}$1$}, %
        {\texttt{qGRAND-}$10^{7}$},
    }

     \end{loglogaxis}
\end{tikzpicture}}\label{fig:89_LER}}\hspace{-1.5mm}\hfill
    \subfloat[]{\makebox[0.245\linewidth][r]{\definecolor{curvegray}{RGB}{125,125,125}
\definecolor{AEDorange}{RGB}{235,85,25}
\definecolor{AEDpurple}{RGB}{145,70,180}
\definecolor{AEDteal}{RGB}{0,155,150}
\definecolor{qblue}{RGB}{0,75,215}
\definecolor{newAEDgreen}{RGB}{0,135,85}
\definecolor{newAEDblue}{RGB}{20,110,175}
\definecolor{newAEDred}{RGB}{190,55,55}

\begin{tikzpicture}
\begin{loglogaxis}[
    xlabel={$p$},
    ylabel={ $\eta$
    },
    xmin=1e-3,
    xmax=1e-1,
    ymin=1e4,
    ymax=5e7,
    grid=both,
    major grid style={solid,gray!70},
    minor grid style={solid,gray!30},
    height=7.5cm,
    width=\columnwidth,
    legend cell align=left,
    legend pos = south east,
    legend style={legend columns=1, font=\tiny, inner sep=1pt, row sep=-1pt, fill opacity=0.85, text opacity=1},
    tri cost axis,
]

\addplot[
    color= RedOrange, %
    very thick
] coordinates {
    (0.001,4.87133e+06)
    (0.100,4.87133e+06)
};
\aedentry{\texttt{AED-}$8+$\texttt{OSD-}$1$ (full evaluation)}{full}

\addplot[cost noleg, color=ForestGreen,mark=*,mark size = 2.5pt,thick, mark options = {solid}] coordinates {
    (0.001,723707) (0.002,991346) (0.003,1236229)
    (0.004,1464190) (0.005,1655297) (0.006,1845561)
    (0.007,2056456) (0.008,2182088) (0.009,2316526)
    (0.010,2426748) (0.020,3196682) (0.030,3609186)
    (0.040,3736495) (0.050,3772228) (0.060,3803533)
    (0.070,3803533) (0.080,3803533) (0.090,3803533)
    (0.100,3803533)
};
\costentry{\texttt{BP}$2+$\texttt{OSD-}$60$}

\addplot[
    color= RedOrange, %
    mark=triangle*, densely dotted,
    mark size=2.5pt,    
    thick,
    mark options={solid}
] coordinates {
 (0.001,701580)
    (0.002,943597)
    (0.003,1.16625e+06)
    (0.004,1.37109e+06)
    (0.005,1.55956e+06)
    (0.006,1.73296e+06)
    (0.007,1.89251e+06)
    (0.008,2.03932e+06)
    (0.009,2.1744e+06)
    (0.01,2.2987e+06)
    (0.02,3.10775e+06)
    (0.03,3.45903e+06)
    (0.04,3.6106e+06)
    (0.05,3.67533e+06)
    (0.06,3.7026e+06)
    (0.07,3.7139e+06)
    (0.08,3.7185e+06)
    (0.09,3.72033e+06)
    (0.1,3.72104e+06)
};
\aedentry{\texttt{AED-}$8+$\texttt{OSD-}$1$ (progressive)}{progressive}

\addplot[cost noleg, color=ForestGreen,mark=*,densely dashed,mark size = 2.5pt,thick, mark options = {solid}] coordinates {
    (0.001,459613) (0.002,479626) (0.003,497937)
    (0.004,514983) (0.005,529273) (0.006,543500)
    (0.007,559270) (0.008,568664) (0.009,578716)
    (0.010,586958) (0.020,644530) (0.030,675375)
    (0.040,684895) (0.050,687567) (0.060,689908)
    (0.070,689908) (0.080,689908) (0.090,689908)
    (0.100,689908)
};
\costentry{\texttt{BP}$2+$\texttt{OSD-}$10$}

\addplot[ 
        cost noleg,
        color = blue, 
        mark = asterisk, thick, densely dashed,
        mark size = 2.5pt,
        mark options = {solid}
        ]
        coordinates {
    (0.001,18718) (0.002,51632) (0.003,97547)
    (0.004,156385) (0.005,223094) (0.006,303862)
    (0.007,415800) (0.008,457839) (0.009,565274)
    (0.010,688315) (0.020,1650789) (0.030,2021028)
    (0.040,2298241) (0.050,2672395) (0.060,3173565)
    (0.070,2829133) (0.080,2844171) (0.090,2846853)
    (0.100,3058689)
};
\costentry{\texttt{qGRAND-}$10^{7}$}

\end{loglogaxis}

\end{tikzpicture}}\label{fig:costs_89}}\hfill
    \subfloat[]{\makebox[0.245\linewidth][r]{\begin{tikzpicture} 
    \begin{loglogaxis}[
        xlabel={$p$},
        ylabel={LER},
        xmin=0.001, xmax=0.1,
        ymin=0.0000005, ymax=1,
        grid=both,
        major grid style={solid, gray!70},
        minor grid style={solid, gray!30},
        legend pos = south east,
        height  = 7.5cm,
        width = \columnwidth, 
        legend cell align=left,
        legend style={legend columns=1, font=\tiny, inner sep=1pt, row sep=-1pt, fill opacity=0.85, text opacity=1},
        scaled x ticks=false,
        tri axis=leg95
    ]

    \addplot[ 
        color = ForestGreen, 
        mark = o, thick, densely dashed,
        mark size = 2.5pt,
        mark options = {solid},
        ]
        coordinates {
            (0.001, 2.2954e-02)
            (0.002, 4.5851e-02)
            (0.003, 6.8724e-02)
            (0.004, 9.1593e-02)
            (0.005, 1.1446e-01)
            (0.006, 1.3732e-01)
            (0.007, 1.6016e-01)
            (0.008, 1.8296e-01)
            (0.009, 2.0568e-01)
            (0.01, 2.2831e-01)
            (0.02, 4.4232e-01)
            (0.03, 6.1792e-01)
            (0.04, 7.4707e-01)
            (0.05, 8.3536e-01)
            (0.06, 8.9299e-01)
            (0.07, 9.2961e-01)
            (0.08, 9.5259e-01)
            (0.09, 9.6696e-01)
            (0.1, 9.7595e-01)
    };
    \addlegendentry{\texttt{BP}$2$} %

    \addplot[ 
        color = ForestGreen, 
        mark = *, thick, densely dashed,
        mark size = 2.5pt,
        mark options = {solid},
        ]
        coordinates {
            (0.1, 0.4694835680751174)
            (0.09, 0.5154639175257731)
            (0.08, 0.47619047619047616)
            (0.07, 0.4219409282700422)
            (0.06, 0.4166666666666667)
            (0.05, 0.33112582781456956)
            (0.04, 0.24752475247524752)
            (0.03, 0.14792899408284024)
            (0.02, 0.08992805755395683)
            (0.01, 0.021767522855899)
            (0.009, 0.01881821603312006)
            (0.008, 0.015554518587649712)
            (0.007, 0.01038961038961039)
            (0.006, 0.00804246421103426)
            (0.005, 0.005420347986340723)
            (0.004, 0.0037054878274724867)
            (0.003, 0.0019806267806267807)
            (0.002, 0.000985017878074487)
            (0.001, 0.0002574161595568323)
    };
    \addlegendentry{\texttt{BP}$2+$\texttt{OSD-}$10$}

    \addplot[ 
        color = ForestGreen, 
        mark = *, thick,
        mark size = 2.5pt,
        mark options = {solid},
        ]
        coordinates {
            (0.09, 0.47619047619047616)
            (0.08, 0.4484304932735426)
            (0.07, 0.3508771929824561)
            (0.06, 0.4149377593360996)
            (0.05, 0.27472527472527475)
            (0.04, 0.17064846416382254)
            (0.03, 0.11363636363636363)
            (0.02, 0.03544842254519674)
            (0.01, 0.004672897196261682)
            (0.009, 0.002952378140592247)
            (0.008, 0.002541231480775584)
            (0.007, 0.0013012361743656475)
            (0.006, 0.000877108349194376)
            (0.005, 0.00042288483577267405)
            (0.004, 0.00021216602415722352)
            (0.003, 8.471668622210173e-05)
            (0.002, 2.4135741340156345e-05)
            (0.001, 2.6332934292824703e-06)
    };
     \addlegendentry{\texttt{BP}$2+$\texttt{OSD-}$60$}

    \addplot[ 
        color = RedOrange, 
        mark = triangle*, thick, densely dashed,
        mark size = 2.5pt,
        mark options = {solid},
        ]
        coordinates {
            (0.001, 5.9507e-03)
            (0.002, 1.1892e-02)
            (0.003, 1.7944e-02)
            (0.004, 2.4210e-02)
            (0.005, 3.0780e-02)
            (0.006, 3.7725e-02)
            (0.007, 4.5105e-02)
            (0.008, 5.2964e-02)
            (0.009, 6.1335e-02)
            (0.01, 7.0240e-02)
            (0.02, 1.8757e-01)
            (0.03, 3.3686e-01)
            (0.04, 4.8526e-01)
            (0.05, 6.1221e-01)
            (0.06, 7.1133e-01)
            (0.07, 7.8461e-01)
            (0.08, 8.3718e-01)
            (0.09, 8.7438e-01)
            (0.1, 9.0053e-01)
    };
    \addlegendentry{$\texttt{AED-}8$} %

    \addplot[ 
        color = RedOrange, 
        mark = triangle*, thick,
        mark size = 2.5pt,
        mark options = {solid},
        ]
        coordinates {
            (0.001, 4.9476e-07)
            (0.002, 7.5043e-06)
            (0.003, 3.6025e-05)
            (0.004, 1.0800e-04)
            (0.005, 2.5017e-04)
            (0.006, 4.9231e-04)
            (0.007, 8.6579e-04)
            (0.008, 1.4024e-03)
            (0.009, 2.1332e-03)
            (0.01, 3.0883e-03)
            (0.02, 2.9749e-02)
            (0.03, 9.1422e-02)
            (0.04, 1.7644e-01)
            (0.05, 2.6508e-01)
            (0.06, 3.4209e-01)
            (0.07, 4.0070e-01)
            (0.08, 4.4094e-01)
            (0.09, 4.6641e-01)
            (0.1, 4.8160e-01)
    };
    \addlegendentry{$\texttt{AED-}8+$\texttt{OSD-}$1$} %
    
    \addplot[ 
        color = blue, 
        mark = asterisk, thick,
        mark size = 2.5pt,
        mark options = {solid}
        ]
        coordinates {
            (0.0005, 1.2546e-07)
            (0.0006, 2.1192e-07)
            (0.0007, 4.2244e-07)
            (0.0008, 6.3807e-07)
            (0.0009, 1.2689e-06)
            (0.0010, 1.4929e-06)
            (0.0015, 8.4011e-06)
            (0.0020, 2.613e-05)
            (0.0030, 0.00014198)
            (0.0040, 0.00041927)
            (0.0050, 0.00079025)
            (0.0060, 0.0016046)
            (0.0070, 0.0032247)
            (0.0080, 0.0042012)
            (0.0090, 0.0062858)
            (0.0100, 0.0099711)
            (0.0150, 0.038432)
            (0.0200, 0.084459)
            (0.0250, 0.14556)
            (0.0300, 0.22676)
            (0.0400, 0.4386)
            (0.0500, 0.54645)
            (0.0600, 0.66667)
            (0.0700, 0.72464)
            (0.0800, 0.84746)
            (0.0900, 0.90909)
            (0.1000, 0.95238)
    };
    \addlegendentry{$\texttt{qGRAND-}10^{6}$}

    \addplot[ 
        color = blue, 
        mark = asterisk, thick, dashed,
        mark size = 2.5pt,
        mark options = {solid},
        ]
        coordinates {
            (0.0006, 5.200e-08) %
            (0.0007, 1.064e-07)
            (0.0008, 1.704e-07)
            (0.0009, 3.039e-07)
            (0.0010, 5.035e-07)
            (0.0015, 2.2263e-06)
            (0.0020, 7.454e-06)
            (0.0030, 4.1446e-05)
            (0.0040, 0.00010927)
            (0.0050, 0.00028793)
            (0.0060, 0.00056864)
            (0.0070, 0.0010685)
            (0.0080, 0.0020293)
            (0.0090, 0.0026455)
            (0.0100, 0.0037703)
            (0.0150, 0.016263)
            (0.0200, 0.038971)
            (0.0250, 0.08285)
            (0.0300, 0.13193)
            (0.0400, 0.27778)
            (0.0500, 0.49751)
            (0.0600, 0.5291)
            (0.0700, 0.71429)
            (0.0800, 0.72993)
            (0.0900, 0.81967)
            (0.1000, 0.91743)
    };
    \addlegendentry{$\texttt{qGRAND-}10^{7}$}

    \end{loglogaxis}
\end{tikzpicture}}\label{fig:95_LER}}\hspace{-0.5mm}\hfill
    \subfloat[]{\makebox[0.245\linewidth][r]{\definecolor{newAEDgreen}{RGB}{0,135,85}
\definecolor{newAEDpurple}{RGB}{145,70,180}

\begin{tikzpicture}
    \begin{loglogaxis}[
        xlabel={$p$},
        ylabel={$\eta$},
        xmin=0.001,
        xmax=0.1,
        ymin=4e3,
        ymax=1e11,
        grid=both,
        major grid style={solid, gray!70},
        minor grid style={solid, gray!30},
        legend pos = south east,
        height  = 7.5cm,
        width = \columnwidth,
        legend cell align=left,
        legend style = {legend columns=1, font=\tiny, inner sep=1pt, row sep=-1pt, fill opacity=0.85, text opacity=1}, %
        scaled x ticks=false,
        tri cost axis,
        ytick={1e4,1e6,1e8,1e10}
    ]

    \addplot[
        color= RedOrange, %
        very thick
    ] coordinates {
        (0.001,6216136.7221)
        (0.100,6216136.7221)
    };
    \aedentry{\texttt{AED-}$8+$\texttt{OSD-}$1$ (full evaluation)}{full}

    \addplot[
        color=RedOrange, %
        mark=triangle*, densely dotted,
        thick,
        mark size=2.5pt,
        mark options={solid}
    ] coordinates {
        (0.001,633296.695146) (0.002,860439.959722)
        (0.003,1067851.349140) (0.004,1257612.435640)
        (0.005,1431681.409290) (0.006,1591881.255660)
        (0.007,1739893.218680) (0.008,1877254.461760)
        (0.009,2005359.015440) (0.010,2125461.250780)
        (0.020,3076075.245650) (0.030,3852342.426350)
        (0.040,4530063.491980) (0.050,5073900.853880)
        (0.060,5472389.927320) (0.070,5744990.003230)
        (0.080,5922920.087400) (0.090,6035560.964970)
        (0.100,6105348.154670)
    };
    \aedentry{\texttt{AED-}$8+$\texttt{OSD-}$1$ (progressive)}{progressive}

    \addplot[
        cost noleg,
        color=ForestGreen,
        mark=*,
        thick,
        mark size=2.5pt,
        mark options={solid}
    ]
    coordinates {
        (0.1, 7138941)
        (0.09, 7138941)
        (0.08, 7138941)
        (0.07, 7115235)
        (0.06, 7138941)
        (0.05, 7120380)
        (0.04, 7035180)
        (0.03, 6755081)
        (0.02, 6130700)
        (0.01, 4562847)
        (0.009, 4268104)
        (0.008, 3985452)
        (0.007, 3678251)
        (0.006, 3330326)
        (0.005, 2941093)
        (0.004, 2525022)
        (0.003, 2065289)
        (0.002, 1552158)
        (0.001, 995829)
    };
    \costentry{\texttt{BP}$2+$\texttt{OSD-}$60$}

    \addplot[
        cost noleg,
        color=blue,
        mark=asterisk,
        thick,
        mark size = 2.5pt,
        mark options={solid}
    ]
    coordinates {
        (0.0007, 57000)
        (0.0008, 71250)
        (0.0009, 90250)
        (0.001,  109250)
        (0.002,  584250)
        (0.003,  1890500)
        (0.004,  4365250)
        (0.005,  7742500)
        (0.006,  13775000)
        (0.007,  23289250)
        (0.008,  29972500)
        (0.009,  45429000)
        (0.010,  68666000)
        (0.020,  465186500)
        (0.030,  1151922500)
        (0.040,  2060502500)
        (0.050,  2584427500)
        (0.060,  3185635000)
        (0.070,  3192522500)
        (0.080,  4000450000)
        (0.090,  4377932500)
        (0.100,  4310245000)
    };
    \costentry{\texttt{qGRAND-}$10^{6}$}

    \addplot[
        cost noleg,
        color=blue,
        mark=asterisk,
        thick,
        densely dashed,
        mark size = 2.5pt,
        mark options={solid}
    ]
    coordinates {
        (0.0007, 57000)
        (0.0008, 76000)
        (0.0009, 95000)
        (0.001,  118750)
        (0.002,  693500)
        (0.003,  2569750)
        (0.004,  6683250)
        (0.005,  13646750)
        (0.006,  25056250)
        (0.007,  47105750)
        (0.008,  69055500)
        (0.009,  107625500)
        (0.010,  151710250)
        (0.020,  1505797500)
        (0.030,  4743160000)
        (0.040,  11126875000)
        (0.050,  17939325000)
        (0.060,  22757250000)
        (0.070,  32246325000)
        (0.080,  33545450000)
        (0.090,  36166500000)
        (0.100,  39991675000)
    };
    \costentry{\texttt{qGRAND-}$10^{7}$}

    \end{loglogaxis}
\end{tikzpicture}}\label{fig:costs_95}\hspace{-4.6mm}}\\[0.5mm]
    \pgfplotslegendfromname{leg95}
  \end{minipage}
  \caption{(a) and (c) \acs{LER} and (b) and (d) estimated average number of binary operations $\eta$ per decoded frame for the $\llbracket 89,1,5\rrbracket$ and  $\llbracket 95,1,7\rrbracket$ triorthogonal codes, respectively, on the dephasing channel. In (b) and (d), the solid and dotted orange \texttt{AED-}$8+$\texttt{OSD-}$1$ curves correspond to full and progressive evaluation, respectively.}
  \label{fig:Tri_89}\label{fig:Tri95}
\end{figure*}
}

In this section, we describe the decoding methods and evaluate the performance of the considered triorthogonal codes.

\subsection{Decoding Methods}
\label{subsec:decoders}

We consider five decoding strategies adapted to the quantum \ac{CSS} setting:
binary \ac{BP} (\texttt{BP}$2$), 
binary \ac{BP} plus \ac{OSD} post-processing with search order $\lambda$ (\texttt{BP$2+$OSD-}$\lambda$),
BP-based AED using $n_{\tn{AED}}$ nontrivial automorphisms (\texttt{AED-}$n_{\tn{AED}}$), AED with \ac{OSD} post-processing with search order $1$ (\texttt{AED-}$n_{\tn{AED}}+$\texttt{OSD-}$1$), and \ac{GRAND}-based decoding (\texttt{qGRAND-Max\_Query}).

All BP-based decoders use the normalized \ac{MS} algorithm with a maximum of $30$ iterations. We use empirically optimized scaling factors $\alpha=0.75$ for BP-based decoders without \ac{OSD} and $\alpha=0.05$ for those with \ac{OSD}.

For the \texttt{BP}$2+$\texttt{OSD-}$\lambda$ decoder, we use the implementation of~\cite{RoffeWhiteBurtonCampbell20_1}. A failed BP decoding attempt is followed by OSD-$0$ and then by a combination sweep (CS) search. The parameter $\lambda\in\{10,60\}$ controls the CS search depth. Although triorthogonal codes do not generally have a sparse \ac{PCM}, we adopt this decoding strategy because the OSD post-processing stage remains effective beyond the strictly sparse regime.

The \texttt{AED-}$n_{\tn{AED}}$ decoder uses $n_{\tn{AED}}$ nontrivial automorphisms of the classical code $\code{C}_{\tn{X}}$ to generate transformed syndromes, each of which is decoded by BP~\cite{KoutsioumpasSayginelWebsterBrowne25_1sub}. The resulting error estimates are mapped back to the original coordinate ordering. We also consider \texttt{AED-}$n_{\tn{AED}}+$\texttt{OSD-}$1$, which applies \ac{OSD} post-processing with the CS method, using $\lambda = 1$, to the soft output of each \ac{BP} decoding path.
We compare the full and progressive evaluation of the ensemble. Full evaluation runs all \ac{BP} paths and, when enabled, all \ac{OSD} stages on every frame\ifthenelse{\boolean{arXiv}}{, and outputs the minimum-weight syndrome-consistent error estimate among all of them}{}. Progressive evaluation processes the \ac{BP} paths sequentially and stops when a syndrome-consistent error estimate of weight at most $t_{\tn{X}}=\lfloor(d_{\tn{X}}-1)/2\rfloor$ is found. 
For \texttt{AED-}$n_{\tn{AED}}+$\texttt{OSD-}$1$, unresolved frames proceed to \ac{OSD}-$0$ and then \ac{OSD}-$1$, with the same stopping rule after each path. The full ensemble can be evaluated in parallel, but our complexity measure counts the total number of binary operations per frame rather than decoding latency, so that progressive \ac{AED} is more favorable.

Finally, \texttt{qGRAND-Max\_Query} is based on \ac{GRAND}. The syndromes for the \texttt{Max\_Query} most likely error candidates are calculated sequentially, starting with the most likely candidate.
This method is well-suited to medium- and high-rate classical codes and does not require a particular algebraic structure or a sparse \ac{PCM}~\cite{DuffyLiMedard19_1}.

\subsection{Decoding Performance and Complexity}
\label{subsec:disc}

We consider two $\mat{X}$-degenerate triorthogonal codes $\llbracket 89,1,5\rrbracket$ and $\llbracket 95,1,7\rrbracket$. 
The first is a new code obtained with the proposed \ac{ILP} construction (see Table~\ref{tab:code-parameters_nontrivial-TriOrthogonal-codes}), while the second was obtained in~\cite{Sullivan24_1, JainAlbert25_1} through the doubling construction of~\cite{BravyiCross15_1sub}. We assess their \ac{LER} performance using Monte Carlo simulations over the dephasing channel. 
Each simulation point is obtained by simulating until $100$ logical errors are observed. We denote by $\eta$ the
estimated 
average number of binary operations per decoded frame, derived in \ifthenelse{\boolean{arXiv}}{Appendix~\ref{sec:complex_estimation}}{\cite[Appendix~B]{BaldelliMostadLinRosnesBattaglioni26_1sub}}.

For \ac{AED}, we use permutation automorphisms of the classical code $\code{C}_{\tn{X}}$, i.e., coordinate permutations $\pi$ with $\pi(\code{C}_{\tn{X}})=\code{C}_{\tn{X}}$.
We obtain them from the code structure. 
For the $\llbracket 89,1,5\rrbracket$ code, $\code{C}_{\tn{X}}^{\perp}$ is cyclic; indexing its coordinates by $\Integers_{89}$, it is invariant under $u\mapsto u+1$ and $u\mapsto 2u$~\cite{MacWilliamsSloane77_1}, which generate a group of order $89\cdot 11=979$. 
The $\llbracket 95,1,7\rrbracket$ code is a doubled code~\cite{BravyiCross15_1sub, JainAlbert25_1}: its qubits split into two copies of the coordinates of the $[23,11,8]$ even-weight Golay subcode $\code{C}_{\tn{g}}$ and the $49$ qubits of the $\llbracket 49,1,5\rrbracket$ code~\cite{BravyiHaah12_1}. 
Its automorphisms include the automorphisms of $\code{C}_{\tn{g}}$ applied to both copies, swaps of the two copies on the support of any codeword of $\code{C}_{\tn{g}}$, and the automorphisms of the $\llbracket 49,1,5\rrbracket$ code on its $49$ qubits. 
We use only $n_{\tn{AED}}$ of these automorphisms, drawn at random from the groups above. In particular, $n_{\tn{AED}}=8$ automorphisms suffice to achieve the best performance for both the $\llbracket 89,1,5\rrbracket$ and $\llbracket 95,1,7\rrbracket$ codes.

In Fig.~\ref{fig:89_LER}, we evaluate the performance of the novel $\llbracket 89,1,5\rrbracket$ triorthogonal code.
The BP-based decoders \texttt{BP}$2$ (dashed green curve) and \texttt{AED-}$8$ (dashed orange curve) show poor performance, consistent with the nonsparse structure of the \ac{PCM}. 
However, the OSD post-processing stage substantially improves the performance, as shown by \texttt{BP}$2+$\texttt{OSD-}$10$ and \texttt{BP}$2+$\texttt{OSD-}$60$ (dashed and solid green curves with filled markers, respectively). The best overall performance is achieved by \texttt{AED-}$8+$\texttt{OSD-}$1$ and \texttt{qGRAND-}$10^{7}$ (solid orange and dashed blue curves, respectively). 
In particular, Fig.~\ref{fig:costs_89} shows that \texttt{qGRAND-}$10^{7}$ requires fewer average binary operations than the progressive \texttt{AED-}$8+$\texttt{OSD-}$1$ over the considered range of $p$. The latter has about the same cost as \texttt{BP}$2+$\texttt{OSD-}$60$ but a lower \ac{LER}.

Moreover, Fig.~\ref{fig:95_LER} reports the performance of the \(\llbracket 95,1,7 \rrbracket\)
triorthogonal code~\cite{Sullivan24_1, JainAlbert25_1}. This code is \(\mat X\)-degenerate, with
\(d_{\tn X}=7\) and \(d_0=4\).
For this triorthogonal instance, \texttt{BP}$2$, \texttt{AED-}$8$,
and \texttt{BP}$2+$\texttt{OSD-}$10$ show poor performance, consistent with the nonsparse structure of its \ac{PCM}. 
On the other hand, \texttt{BP}$2+$\texttt{OSD-}$60$ substantially improves the performance as a result of the larger OSD search depth. 
The best overall performance is achieved by \texttt{qGRAND-}$10^{7}$ and \texttt{AED-}$8+$\texttt{OSD-}$1$.
Reducing the query limit to $10^{6}$ (\texttt{qGRAND-}$10^{6}$, solid blue curve) increases the \ac{LER} by a factor of about $3$ at the smallest values of $p$, while it lowers the average number of binary operations substantially only for larger $p$. As shown in Fig.~\ref{fig:costs_95}, \texttt{qGRAND-}$10^{7}$ requires fewer average binary operations at the smallest values of $p$, whereas the progressive evaluation of \texttt{AED-}$8+$\texttt{OSD-}$1$ becomes less costly from $p=3\cdot10^{-3}$ onward among the simulated points.

For \texttt{qGRAND-}$10^{7}$, the average computational cost is substantially larger for the $\llbracket 95,1,7\rrbracket$ code than for the $\llbracket 89,1,5\rrbracket$ code. This difference reflects both the larger syndrome dimension, with $\rk(\mat{H}_{\tn{X}})=25$ instead of $11$, and the larger average number of guesses required by \texttt{qGRAND}. Indeed, as detailed in \ifthenelse{\boolean{arXiv}}{Appendix~\ref{subsec:qGRAND}}{~\cite[Appendix~B-C]{BaldelliMostadLinRosnesBattaglioni26_1sub}}, the estimated cost is proportional to $2rn$ times the average number of guesses per frame.

\section{Conclusion}
\label{sec:conclusions}

This paper studied the construction and decoding of triorthogonal codes. We proposed an \ac{ILP} formulation for constructing triorthogonal matrices subject to dual-distance requirements, combining triorthogonality and distance constraints within the same framework. 
In this way, we introduced novel nontrivial degenerate and non-degenerate triorthogonal codes. \ifthenelse{\boolean{arXiv}}{With the automorphism group of the known $\llbracket 49,1,5\rrbracket$ or $\llbracket 95,1,7\rrbracket$ code prescribed, the \ac{ILP} returns exactly that code.}{} The numerical results over the dephasing channel show that \texttt{AED-}$n_{\tn{AED}}+$\texttt{OSD-}$1$ and \texttt{qGRAND-}$10^{7}$ are promising decoding options, achieving good \ac{LER} performance with competitive computational complexity.

\section*{Statement on the Use of AI Tools}
The authors wrote the MATLAB code used to construct the triorthogonal codes in this paper and to verify their triorthogonality and minimum distances. Claude (Anthropic) was used to speed up the verification and reproduction of the reported results, and to revise this MATLAB code. The authors independently checked all the results included in the paper and take full responsibility for its content.

\ifthenelse{\boolean{arXiv}}{%

\appendices

\section{Proof of Theorem~\ref{thm:construction_TriOrthogonal_codes_dperp}}
\label{sec:proof_ILP-theorem}

The proof follows the column-multiplicity approach of~\cite[Lemma~23.3.2 and Th.~23.3.7]{HuffmanKimSole21_1}, with additional constraints for even-weight triorthogonality and the target dual distance. We first prove the necessity and then the sufficiency.

\begin{description}[style=unboxed,leftmargin=0pt,labelsep=0.5em,listparindent=\parindent]
\item[Necessity.] Suppose that there exist a binary $[n,k_0]$ code $\code{C}$ with $d(\dual{\code{C}})\geq\dual{d}$ and a full-rank triorthogonal GM $\mat{H}_{\tn{X}}$ of $\code{C}$ having only even-weight rows. Since $d(\dual{\code{C}})\geq\dual{d}\geq 2$, $\mat{H}_{\tn{X}}$ has no zero column, as such a column would give a dual codeword of weight one. Thus, every column of $\mat{H}_{\tn{X}}$ belongs to $\set{V}$. For each $i\in[N]$, let $x_i$ be the number of columns equal to $\vect{v}_i$. Then, $x_i\in\{0\}\cup[n]$ and $\sum_{i=1}^{N}x_i=n$, which verifies $(\tn{L1})$. Moreover, by~\eqref{eq:Pk_self-orthogonality} and~\eqref{eq:Tk_tri-orthogonality}, we have
  \begin{IEEEeqnarray*}{rCl}
    {}[\mat{P}_{k_0}\vect{x}]_{(a,b)}& = &\sum_{j=1}^{n}[\mat{H}_{\tn{X}}]_{a,j}[\mat{H}_{\tn{X}}]_{b,j},
    \\
    {}[\mat{T}_{k_0}\vect{x}]_{(a,b,c)}& = &\sum_{j=1}^{n}[\mat{H}_{\tn{X}}]_{a,j}[\mat{H}_{\tn{X}}]_{b,j}[\mat{H}_{\tn{X}}]_{c,j}.
  \end{IEEEeqnarray*}
 
  For $a=b$, the first expression is the weight of the $a$-th row of $\mat{H}_{\tn{X}}$, which is even by assumption. For $a<b$, the first expression is the pairwise overlap of rows $a$ and $b$, and the second expression is the triple overlap of rows $a$, $b$, and $c$. Both are even by Definition~\ref{def:TriOrthogonal-GM} and Remark~\ref{rem:even-overlap_Hx}. Hence, $\vect{z}_{\tn{P}}=\mat{P}_{k_0}\vect{x}/2$ and $\vect{z}_{\tn{T}}=\mat{T}_{k_0}\vect{x}/2$ have nonnegative integer entries and satisfy $(\tn{O1})$ and $(\tn{O2})$.

  For each $i\in[N]$, a column of type $\vect{v}_{i'}$ contributes a zero entry to the codeword $\trans{\mat{H}_{\tn{X}}}\vect{v}_i$ if and only if $\einner{\vect{v}_i}{\vect{v}_{i'}}=0$, i.e., if and only if $[\mat{M}]_{i,i'}=1$. Since $x_{i'}$ is the number of columns of type $\vect{v}_{i'}$, the quantity $[\mat{M}\vect{x}]_i=\sum_{i'=1}^{N}[\mat{M}]_{i,i'}x_{i'}$ counts the zero positions in $\trans{\mat{H}_{\tn{X}}}\vect{v}_i$. Therefore,
  \begin{equation}
    \wt(\trans{\mat{H}_{\tn{X}}}\vect{v}_i)=n-[\mat{M}\vect{x}]_i.
    \label{eq:proof_ILP_weight}
  \end{equation}
  Since $\mat{H}_{\tn{X}}$ has rank $k_0$, the vectors $\trans{\mat{H}_{\tn{X}}}\vect{v}_i$, $i\in[N]$, are precisely the distinct nonzero codewords of $\code{C}$. Set $\delta_{i,j}=1$ if $\wt(\trans{\mat{H}_{\tn{X}}}\vect{v}_i)=j$ and $\delta_{i,j}=0$ otherwise. Since these weights belong to $[n]$, $(\tn{W1})$ holds, and~\eqref{eq:proof_ILP_weight} gives $(\tn{W2})$.
  
  Let $A_j$ and $B_\ell$ denote the weight enumerator coefficients of $\code{C}$ and $\dual{\code{C}}$, respectively. We have $A_0=1$ and $A_j=\sum_{i=1}^{N}\delta_{i,j}$ for $j\in[n]$. Applying~\eqref{eq:MacWilliams-Theorem} with $\ecard{\code{C}}=2^{k_0}$ and $\mathrm{K}_\ell(0;n)=\binom{n}{\ell}$ gives
  \begin{equation}
    2^{k_0}B_\ell=\binom{n}{\ell}+\sum_{i=1}^{N}\sum_{j=1}^{n}\mathrm{K}_\ell(j;n)\delta_{i,j}.
    \label{eq:proof_ILP_dual_weights}
  \end{equation}
  Since $d(\dual{\code{C}})\geq\dual{d}$, we have $B_\ell=0$ for $\ell\in[\dual{d}-1]$, and hence~\eqref{eq:proof_ILP_dual_weights} gives $(\tn{D=})$. Therefore, the ILP is feasible.

\item[Sufficiency.] Conversely, suppose that the ILP has a feasible solution. Construct $\mat{H}_{\tn{X}}\in\Field_2^{k_0\times n}$ by arranging $x_i$ copies of each column type $\vect{v}_i$, $i\in[N]$, in an arbitrary order; by $(\tn{L1})$, it has $n$ columns. The overlap expressions above also hold for this matrix. Thus, $(\tn{O1})$ guarantees that its row weights and pairwise overlaps are even, and $(\tn{O2})$ guarantees that its triple overlaps are even. Hence, $\mat{H}_{\tn{X}}$ is triorthogonal and has only even-weight rows.
  
  We next show that the $k_0$ rows of $\mat{H}_{\tn{X}}$ are linearly independent. For each $i\in[N]$, $(\tn{W1})$ and the binary constraints imply that exactly one indicator, say $\delta_{i,j_i}$ with $j_i\in[n]$, equals one. Since~\eqref{eq:proof_ILP_weight} holds for any matrix built from $\vect{x}$, $(\tn{W2})$ gives
  \begin{equation*}
    \wt(\trans{\mat{H}_{\tn{X}}}\vect{v}_i)=n-[\mat{M}\vect{x}]_i=\sum_{j=1}^{n}j\delta_{i,j}=j_i\geq 1.
  \end{equation*}
  As $\vect{v}_i$ ranges over all nonzero vectors in $\Field_2^{k_0}$, every nontrivial linear combination $\trans{\mat{H}_{\tn{X}}}\vect{v}_i$ of the rows of $\mat{H}_{\tn{X}}$ is nonzero. Hence, the rows are linearly independent, i.e., $\rk(\mat{H}_{\tn{X}})=k_0$, and $\code{C}\eqdef\Rspan(\trans{\mat{H}_{\tn{X}}})$ is a binary $[n,k_0]$ code with full-rank GM $\mat{H}_{\tn{X}}$.
  
  Finally, the vectors $\trans{\mat{H}_{\tn{X}}}\vect{v}_i$, $i\in[N]$, are the distinct nonzero codewords of $\code{C}$, and $(\tn{W2})$ ensures that their weights are those specified by the indicators. Thus, $A_0=1$ and $A_j=\sum_{i=1}^{N}\delta_{i,j}$ are the weight enumerator coefficients of $\code{C}$, and~\eqref{eq:MacWilliams-Theorem} again gives~\eqref{eq:proof_ILP_dual_weights}. By $(\tn{D=})$, $B_\ell=0$ for every $\ell\in[\dual{d}-1]$, and therefore $d(\dual{\code{C}})\geq\dual{d}$, which completes the proof. Note that the proof holds for both even and odd code lengths $n$. Odd length is needed only when the all-ones row is appended to $\mat{H}_{\tn{X}}$ to obtain an $\llbracket n,1,d\rrbracket$ triorthogonal code.
\end{description}

\section{Computational Complexity Estimation}
\label{sec:complex_estimation}

In this appendix, we show in detail how we compute the average estimated number of binary operations per decoded frame for \texttt{BP}$2$, \texttt{BP}$2+$\texttt{OSD-}$\lambda$, \texttt{qGRAND-Max\_Query}, \texttt{AED-}$n_{\tn{AED}}$, and \texttt{AED-}$n_{\tn{AED}}+$\texttt{OSD-}$\lambda$, that are depicted in Fig.~\ref{fig:costs_89} and Fig.~\ref{fig:costs_95}.

\subsection{\texttt{BP}$2+$\texttt{OSD-}$\lambda$ Decoder} 
\label{subsec:BP_OSD}
We account for a $q$-bit quantized implementation, so reliability-domain arithmetic can be counted in terms of binary-operation equivalents.
Moreover, we set
\begin{itemize}
    \item $r = \rk(\mat{H}_{\tn{X}})$ as the binary rank of the considered classical \ac{PCM}, which equals $k_0$ for the full-rank matrices $\mat{H}_{\tn{X}}$ of Section~\ref{sec:main_result},
    \item $k_{\tn{X}} = n - \rk(\mat{H}_{\tn{X}})$ as the dimension of the considered classical code,     
    \item $n_{\tn{iter}}$ as the maximum number of BP iterations,
    \item $n_{\tn{MC}}$ as the number of transmitted frames, or Monte Carlo samples, simulated at a fixed value of the dephasing probability $p$,
    \item $n_{\tn{OSD}}$ as the number of times (for a given value of $p$) for which the post-processing routine is called (both \texttt{OSD-0} and \texttt{CS-$\lambda$}), and
    \item $n_{\tn{e}}$ as the number of edges in the Tanner graph of $\mat{H}_{\tn{X}}$.
\end{itemize}

According to \cite{RoffeWhiteBurtonCampbell20_1}, whenever the BP decoder fails after $n_{\tn{iter}}$ iterations, the corresponding \ac{OSD} post-processing stage is invoked, namely 
\texttt{OSD-}$0$ followed, if needed, by \texttt{CS-}$\lambda$. %
Therefore, $n_{\tn{OSD}}$ can be obtained from a \texttt{BP}$2+$\texttt{OSD-}$\lambda$ run by counting the number of such \ac{BP} failures.

In the following, we denote by \(C(\texttt{BP}2)\) the cost of the BP decoding stage only, implemented through the \ac{MS} check-node update rule. We use a worst-case estimate, assuming that all \(n_{\textnormal{iter}}\) iterations are performed for every frame. This is in line with what we observed for the specific codes we have simulated. It would also be
consistent with removing the early stopping criterion provided by intermediate syndrome checks. The resulting cost is
\begin{equation*}
    C(\texttt{BP}2) = n_{\tn{iter}}(4 q n_{\tn{e}} + n_{\tn{e}} + n).
\end{equation*}
The term $4qn_{\tn{e}}$ accounts for the reliability-domain operations associated with the message updates along the edges. For each edge, we count a constant number of $q$-bit operations, including the update of variable-to-check and check-to-variable messages, reliability additions/subtractions, and minimum/comparison operations. The term $n_{\tn{e}}$ accounts for the binary sign processing in the check-node updates, namely the propagation and combination of message signs along the edges. Finally, the term $n$ accounts for the tentative hard decision on the $n$ variable nodes, where a sign test is performed on each variable node to obtain the current binary estimate. Since these operations are performed at each iteration, the per-iteration cost is multiplied by $n_{\tn{iter}}$.

In the following, we denote by $C(\texttt{OSD-}0)$ the cost of the \texttt{OSD-}$0$ post-processing routine, after the BP decoding stage. 
In particular, according to \cite{RoffeWhiteBurtonCampbell20_1}, we first need to sort the soft-decision vector, i.e., the output of the BP decoding stage, and then re-order the columns of $\mat{H}_{\tn{X}}$ accordingly, obtaining $\mat{H}_{\tn{X,s}} \in \mathbb{F}_2^{r \times n}$, where the subscript ``$\tn{s}$'' stands for ``sorted''.
The corresponding cost is 
\begin{equation*}
  C(\texttt{sort\_OSD})  = q (n \log_2(n)) + n.
\end{equation*}
Next, we have to perform Gaussian elimination in order to find the first $r$ linearly independent columns of $\mat{H}_{\tn{X,s}}$, which has an estimated cost of
\begin{equation*}
    C(\texttt{GE}) = r (r-1) (3 n - r + 2) / 6.
\end{equation*}
In this way, we obtain a full-rank matrix $\tilde{\mat{H}}_{\tn{X,s}} \in \mathbb{F}_2^{r \times r}$, and sorting the columns further we set, without loss of generality,
\begin{equation*}
    \mat{H}_{\tn{X,s}} = (\tilde{\mat{H}}_{\tn{X,s}} \; \mat{H}_{\tn{X,rem}}),
\end{equation*}
where $\mat{H}_{\tn{X,rem}} \in \mathbb{F}_2^{r \times k_{\tn{X}}}$ and the subscript ``$\tn{rem}$'' stands for ``remainder''.

Now, we can rewrite the error vector as
\begin{equation*}
    \bm{e}_{\tn{Z}} = 
    \begin{pmatrix}
        \bm{e}_{\tn{Z,s}} \\ 
        \bm{e}_{\tn{Z,rem}}
    \end{pmatrix}
    \in \mathbb{F}_2^{n},
\end{equation*}
where $\bm{e}_{\tn{Z,s}} \in \mathbb{F}_2^{r}$ and $\bm{e}_{\tn{Z,rem}} \in \mathbb{F}_2^{k_{\tn{X}}}$.
As described in~\cite{RoffeWhiteBurtonCampbell20_1}, the treatment of $\bm{e}_{\tn{Z,rem}}$ depends on the post-processing routine. In general, for a given choice of $\bm{e}_{\tn{Z,rem}}$, the corresponding $\bm{e}_{\tn{Z,s}}$ can be chosen so that the syndrome equation $\vect{s}_{\tn{X}}=\mat{H}_{\tn{X,s}}\vect{e}_{\tn{Z}}$ is satisfied. 
However, for \texttt{OSD-}$0$, $\vect{e}_{\tn{Z,rem}}$ is fixed to the all-zero vector, so that $\bm{e}_{\tn{Z,s}} = \tilde{\mat{H}}_{\tn{X,s}}^{-1}\bm{s}_{\tn{X}}$. Hence, no additional cost is counted for computing $\vect{e}_{\tn{Z,rem}}$.
In contrast, in the second post-processing stage, namely \texttt{CS-}$\lambda$, nonzero configurations of $\vect{e}_{\tn{Z,rem}}$ are considered, and the corresponding cost will be accounted for later on.

In order to find ${\tilde{\mat{H}}_{\tn{X,s}}}^{-1}$, we need to invert $\tilde{\mat{H}}_{\tn{X,s}}$, which has a cost of
\begin{equation*}
  C(\texttt{INV}) = r (r-1)(3 r + 1) / 2,
\end{equation*}
and the cost for computing the product between the syndrome $\bm{s}_{\tn{X}} \in \mathbb{F}_2^{r}$ and $\tilde{\mat{H}}_{\tn{X,s}}^{-1}$, namely $\bm{e}_{\tn{Z,s}} = \tilde{\mat{H}}_{\tn{X,s}}^{-1} \bm{s}_{\tn{X}}$, is
\begin{equation*}
  C(\texttt{prod\_OSD})= r (2r - 1).
\end{equation*}

In summary, the total cost for the \texttt{OSD-}$0$ routine becomes
\begin{IEEEeqnarray*}{rCl}
  C(\texttt{OSD-}0) & = &C(\texttt{sort\_OSD}) 
  \nonumber\\
  && +\> C(\texttt{GE})+ C(\texttt{INV}) + C(\texttt{prod\_OSD}).
\end{IEEEeqnarray*}

Next, let $C(\texttt{CS-$\lambda$})$ denote the cost for the second post-processing routine, after \texttt{OSD-}$0$. According to~\cite{RoffeWhiteBurtonCampbell20_1}, we first sort the entries of the remainder component $\bm{e}_{\tn{Z,rem}}$ according to the \ac{BP} soft-decision reliabilities. This has a cost of
\begin{equation*}
  C(\texttt{sorting\_CS}) = k_{\tn{X}}.
\end{equation*}
Then, we count the possible configurations considered by the \texttt{CS} method. In \texttt{CS}-$\lambda$, we consider all weight-one configurations of $\vect{e}_{\tn{Z,rem}}$, together with all weight-two configurations supported on the first $\lambda$ entries of $\vect{e}_{\tn{Z,rem}}$, where these entries are ordered according to the \ac{BP} soft-decision reliabilities. Therefore, the total number of configurations is
\begin{equation*}
  n_{\tn{conf}} = \binom{k_{\tn{X}}}{1} + \binom{\lambda}{2}= k_{\tn{X}} + \binom{\lambda}{2}.
\end{equation*}
For each of these configurations, we have to compute
\begin{equation} 
  \label{eq:operations}
  \bm{e}_{\tn{Z,s}} = \tilde{\mat{H}}_{\tn{X,s}}^{-1} \bm{s}_{\tn{X}} \, + \, \tilde{\mat{H}}_{\tn{X,s}}^{-1} \mat{H}_{\tn{X,rem}} \bm{e}_{\tn{Z,rem}}.
\end{equation}
The product $\tilde{\mat{H}}_{\tn{X,s}}^{-1} \bm{s}_{\tn{X}}$ is known from \texttt{OSD-}$0$, and the product $\tilde{\mat{H}}_{\tn{X,s}}^{-1} \mat{H}_{\tn{X,rem}}$ needs to be computed only once ($\texttt{precomp}$), while the multiplication with $\bm{e}_{\tn{Z,rem}}$ and the summation must be calculated $n_{\tn{conf}}$ times. Hence, for $\texttt{precomp}$, we have a cost of
\begin{equation*}
  C(\texttt{pre}\texttt{comp\_CS}) = r^2 k_{\tn{X}} + r k_{\tn{X}} (r - 1) =  r k_{\tn{X}} (2r-1),
\end{equation*}
and the total cost of~\eqref{eq:operations} is
\begin{IEEEeqnarray*}{rCl}
  C(\texttt{op}\texttt{erations}) & = & C(\texttt{precomp\_CS}) 
  \nonumber\\
  &&+\> n_{\tn{conf}}(r k_{\tn{X}} + r   (k_{\tn{X}} - 1) + r).
\end{IEEEeqnarray*}
The routine ends after having found the lowest-weight error pattern among all the configurations; hence, we have an additional cost of 
\begin{equation*}
  C(\texttt{comparisons}) = n_{\tn{conf}}   (n - 1) + (n_{\tn{conf}} - 1),
\end{equation*}
where $n_{\tn{conf}} (n - 1)$ is the cost to compute the Hamming weight for each error pattern and $(n_{\tn{conf}} - 1)$ is the cost of the comparison between all the Hamming weights of these vectors, in order to find the lowest-weight vector. 

In summary, the total cost of \texttt{CS-$\lambda$} becomes
\begin{IEEEeqnarray*}{rCl}
  C(\texttt{CS-$\lambda$}) & = &C(\texttt{sorting\_CS}) 
  \nonumber\\
  &&+\> C(\texttt{operations}) + C(\texttt{comparisons}),
\end{IEEEeqnarray*}
and the total cost of the \texttt{OSD-$\lambda$} post-processing routine becomes
\begin{equation*}
  n_{\tn{OSD}}[C(\texttt{OSD-}0) + C(\texttt{CS-$\lambda$})].
\end{equation*}

Finally, the average cost per decoded (correctly decoded or not) frame becomes
\begin{IEEEeqnarray*}{rCl}
  \IEEEeqnarraymulticol{3}{l}{
    C(\texttt{BP}2+\texttt{OSD-$\lambda$})
  }\nonumber\\*\quad
  & = &\frac{n_{\tn{MC}}    C(\texttt{BP}2) + n_{\tn{OSD}} [C(\texttt{OSD-}0) + C(\texttt{CS-$\lambda$})]}{n_{\tn{MC}}}.
\end{IEEEeqnarray*}

\subsection{\texttt{AED-}$n_{\tn{AED}}+$\texttt{OSD-}$\lambda$ Decoder}
\label{subsec:AED_OSD}

We now extend the previous complexity estimate to AED. Let $n_{\tn{AED}}$ denote the number of nontrivial code automorphisms employed by the ensemble. For each automorphism $\pi_i$, $i\in[n_{\tn{AED}}]$, let $\mat{Q}_{\pi_i}$ be its permutation matrix. The corresponding syndrome transformation $\mat{A}_{\pi_i}\in\mathbb{F}_2^{r\times r}$ satisfies
\begin{equation*}
  \mat{A}_{\pi_i}\mat{H}_{\tn{X}}=\mat{H}_{\tn{X}}\mat{Q}_{\pi_i}.
\end{equation*}
Hence, if $\bm{s}_{\tn{X}}$ is the measured syndrome, the $i$-th \ac{AED} path applies the \ac{BP} decoder to the transformed syndrome $\mat{A}_{\pi_i}\bm{s}_{\tn{X}}$ and subsequently maps the resulting soft and hard decisions back to the original coordinate ordering.

The automorphisms, their inverse permutations, and the matrices $\mat{A}_{\pi_i}$ are computed once and stored before decoding.
Therefore, their construction is not included in the per-frame complexity. For each \ac{AED} path, the online syndrome transformation is a dense binary matrix-vector product of size $r\times r$, with cost
\begin{equation*}
  C(\texttt{A}_{\pi})=r^2+r(r-1)=r(2r-1).
\end{equation*}
Applying the inverse coordinate permutation is regarded as a memory reindexing operation and is not assigned an additional binary-operation cost.

Each \ac{AED} path uses the same \ac{PCM} $\mat{H}_{\tn{X}}$ and the same quantized \ac{MS} decoder. Consequently, its \ac{BP} cost is the one derived in Appendix~\ref{subsec:BP_OSD}, namely
\begin{equation*}
  C(\texttt{BP}2) = n_{\tn{iter}}\bigl(4qn_{\tn{e}}+n_{\tn{e}}+n\bigr).
\end{equation*}

For progressive evaluation, let $n_{\tn{AED},\tn{BP}}$ denote the total number of \ac{BP} paths actually executed over the $n_{\tn{MC}}$ simulated frames. Similarly, let $n_{\tn{AED},0}$ and $n_{\tn{AED},\tn{CS}}$ denote the total numbers of \texttt{OSD-}$0$ and \texttt{CS-}$\lambda$ invocations, respectively. These quantities account for early stopping after each decoding path. Using the costs derived above, the average estimated cost per decoded frame is
\begin{IEEEeqnarray*}{rCl}
  \IEEEeqnarraymulticol{3}{l}{%
    C_{\tn{prog}}(\texttt{AED-}n_{\tn{AED}}+\texttt{OSD-}\lambda)
  }\nonumber\\*\quad
  & = &
  \frac{n_{\tn{AED},\tn{BP}}}{n_{\tn{MC}}}
  \left[C(\texttt{A}_{\pi})+C(\texttt{BP}2)\right]
  \nonumber\\
  &&+\>
  \frac{n_{\tn{AED},0}}{n_{\tn{MC}}}
  C(\texttt{OSD-}0)+\frac{n_{\tn{AED},\tn{CS}}}{n_{\tn{MC}}}
  C(\texttt{CS-}\lambda).
\end{IEEEeqnarray*}
For \texttt{AED-}$n_{\tn{AED}}+$\texttt{OSD-}$0$, the last term is omitted.

In full evaluation, all $n_{\tn{AED}}$ paths are executed on every frame, including both \ac{OSD} stages. Consequently,
\begin{equation*}
    n_{\tn{AED},\tn{BP}}
    = n_{\tn{AED},0}
    = n_{\tn{AED},\tn{CS}}
    = n_{\tn{MC}}n_{\tn{AED}},
\end{equation*}
and the average estimated cost becomes
\begin{IEEEeqnarray*}{rCl}
  \IEEEeqnarraymulticol{3}{l}{%
    C_{\tn{full}}(\texttt{AED-}n_{\tn{AED}}+\texttt{OSD-}\lambda)
  }\nonumber\\*\quad
  & = &
  n_{\tn{AED}}
  \bigl[
  C(\texttt{A}_{\pi})
  +C(\texttt{BP}2)
  +C(\texttt{OSD-}0)
  +C(\texttt{CS-}\lambda)
  \bigr].
\end{IEEEeqnarray*}
This cost is independent of the dephasing probability $p$.

Progressive evaluation can reduce the computational cost without changing the logical decision. Suppose that one of the \ac{AED} paths produces a valid candidate $\hat{\bm{e}}_{\tn{Z}}$ satisfying $\wt(\hat{\bm{e}}_{\tn{Z}})\leq t_{\tn{X}}$. Any valid candidate $\hat{\bm{e}}_{\tn{Z}}'$ in the other logical class differs from $\hat{\bm{e}}_{\tn{Z}}$ by a nontrivial logical operator and must therefore satisfy
\begin{equation*}
  \wt\bigl(\hat{\bm{e}}_{\tn{Z}}+\hat{\bm{e}}_{\tn{Z}}'\bigr)\geq d_{\tn{X}}.
\end{equation*}
It follows that
\begin{equation*}
  \wt(\hat{\bm{e}}_{\tn{Z}}')
  \geq d_{\tn{X}}-\wt(\hat{\bm{e}}_{\tn{Z}})>t_{\tn{X}}.
\end{equation*}
Since full evaluation outputs the minimum-weight valid candidate, and every valid candidate in the other logical class has weight larger than $t_{\tn{X}}\geq\wt(\hat{\bm{e}}_{\tn{Z}})$, the output of full evaluation lies in the same logical class as $\hat{\bm{e}}_{\tn{Z}}$. If no valid candidate of weight at most $t_{\tn{X}}$ is found, progressive evaluation executes all paths and stages, and the two coincide. Hence, progressive and full evaluation yield the same logical decision and differ only in cost.

In progressive evaluation, the three invocation counts are obtained directly from the decoder by accumulating the number of paths executed at each stage over all simulated frames. The decoder first evaluates the \ac{BP} paths sequentially and, for unresolved frames, proceeds to \texttt{OSD-}$0$ and then \texttt{CS-}$\lambda$, stopping at any stage when a valid error estimate of weight at most $t_{\tn{X}}$ is found. Consequently, the average numbers of invocations depend on the dephasing probability $p$. The resulting complexity is given by $C_{\tn{prog}}$ above, while $C_{\tn{full}}$ describes the full evaluation of the ensemble.

\subsection{\texttt{qGRAND} Decoder} 
\label{subsec:qGRAND}

Here, we particularize the cost for the \ac{GRAND}-based quantum decoder considered in this work, namely \texttt{qGRAND-Max\_Query}. Our implementation is derived from the publicly available MATLAB code associated with~\cite{DuffyLiMedard19_1}, with the modifications required to support the \acp{QSC} and simulation settings considered in this work. In addition to what we already have set in the previous subsection, we denote by $n_{\tn{g}}$ the total number of guesses performed, for a fixed $p$.

For each guessed error pattern, \texttt{qGRAND-Max\_Query} performs the following operations:
\begin{enumerate}
\item it computes the binary syndrome associated with the guessed error pattern and compares it with the measured syndrome, returning the error pattern if the two syndromes are equal;
\item otherwise, it generates the next guessed error pattern, unless \(\texttt{Max\_Query}\) has been reached.
\end{enumerate}
If no valid error pattern is found after \(\texttt{Max\_Query}\) guesses, the decoder stops without any additional cost being included in the present estimate.

For Step 1), the syndrome computation is counted as a dense binary matrix-vector product between \(\mat H_{\tn{X}}\in\mathbb F_2^{r\times n}\) and the guessed error pattern, followed by the comparison with the measured syndrome, and the corresponding cost becomes
\begin{IEEEeqnarray*}{c}
  C(\texttt{syndrome}) = rn+r(n-1)+r.
\end{IEEEeqnarray*}
Step 2) can be done with no computational cost, assuming the potential error patterns are loaded into memory. The space complexity of this is (proportional to) $n\cdot\texttt{Max\_Query}$. In summary, the average estimated cost per decoded frame is therefore
\begin{IEEEeqnarray*}{rCl}    
  \IEEEeqnarraymulticol{3}{l}{
    C(\texttt{qGRAND-}\texttt{Max\_Query}) =\frac{n_{\tn{g}}} {n_{\tn{MC}}} C(\texttt{syndrome})}
  \nonumber\\*\quad%
  & = & \frac{n_{\tn{g}}}{n_{\tn{MC}}} \bigl[rn+r(n-1)+r\bigr] =\frac{n_{\tn{g}}}{n_{\tn{MC}}} 2rn.
\end{IEEEeqnarray*}

\begin{remark}
  In contrast to decoding a classical code, decoding a \ac{QSC} must also account for the presence of degenerate errors. Thus, according to~\cite{RoffeWhiteBurtonCampbell20_1}, after having computed the residual error, we multiply it by the logical operators associated to $\mat{H}_{\tn{X}}$. Then, if the result is the all-zero vector, the decoding procedure is counted as successful. Otherwise, the decoder encounters a failure. Note that, for the \ac{BP}-based decoders and \texttt{qGRAND-Max\_Query}, the computational cost of evaluating the residual error and multiplying it by logical operators is not taken into account. This reflects the practical setting, where we are unable to check at each step if the decoding procedure was successful or not.
\end{remark}
}

\ifthenelse{\boolean{arXiv}}{%
}{
\IEEEtriggeratref{12}
}
\bibliographystyle{IEEEtran}
\bibliography{defshort1, biblioHY, Archive}

\begin{thebibliography}{10}
\providecommand{\url}[1]{#1}
\csname url@samestyle\endcsname
\providecommand{\newblock}{\relax}
\providecommand{\bibinfo}[2]{#2}
\providecommand{\BIBentrySTDinterwordspacing}{\spaceskip=0pt\relax}
\providecommand{\BIBentryALTinterwordstretchfactor}{4}
\providecommand{\BIBentryALTinterwordspacing}{\spaceskip=\fontdimen2\font plus
\BIBentryALTinterwordstretchfactor\fontdimen3\font minus
  \fontdimen4\font\relax}
\providecommand{\BIBforeignlanguage}[2]{{%
\expandafter\ifx\csname l@#1\endcsname\relax
\typeout{** WARNING: IEEEtran.bst: No hyphenation pattern has been}%
\typeout{** loaded for the language `#1'. Using the pattern for}%
\typeout{** the default language instead.}%
\else
\language=\csname l@#1\endcsname
\fi
#2}}
\providecommand{\BIBdecl}{\relax}
\BIBdecl

\bibitem{Gottesman97_1}
D.~Gottesman, ``Stabilizer codes and quantum error correction,'' Ph.D.
  dissertation, California Institute of Technology, 1997.

\bibitem{CalderbankShor96_1}
A.~R. Calderbank and P.~W. Shor, ``Good quantum error-correcting codes exist,''
  \emph{Phys. Rev. A}, vol.~54, no.~2, pp. 1098--1105, Aug. 1996.

\bibitem{Steane96_1}
A.~M. Steane, ``Error correcting codes in quantum theory,'' \emph{Phys. Rev.
  Lett.}, vol.~77, no.~5, pp. 793--797, Jul. 1996.

\bibitem{BravyiKitaev05_1}
S.~Bravyi and A.~Kitaev, ``Universal quantum computation with ideal {Clifford}
  gates and noisy ancillas,'' \emph{Phys. Rev. A}, vol.~71, no.~2, Feb. 2005,
  {Art.} no. 022316.

\bibitem{BravyiHaah12_1}
S.~Bravyi and J.~Haah, ``Magic-state distillation with low overhead,''
  \emph{Phys. Rev. A}, vol.~86, no.~5, Nov. 2012, {Art.} no. 052329.

\bibitem{HaahHastingsPoulinWecker17_1}
J.~Haah, M.~B. Hastings, D.~Poulin, and D.~Wecker, ``Magic state distillation
  with low space overhead and optimal asymptotic input count,'' \emph{Quantum},
  vol.~1, Oct. 2017, {Art.} no. 31.

\bibitem{HaahHastings18_1}
J.~Haah and M.~B. Hastings, ``Codes and protocols for distilling {T},
  controlled-{S}, and {Toffoli} gates,'' \emph{Quantum}, vol.~2, Jun. 2018,
  {Art.} no. 71.

\bibitem{NezamiHaah22_1}
S.~Nezami and J.~Haah, ``Classification of small triorthogonal codes,''
  \emph{Phys. Rev. A}, vol. 106, no.~1, Jul. 2022, {Art.} no. 012437.

\bibitem{JainAlbert25_1}
S.~P. Jain and V.~V. Albert, ``Transversal {Clifford} and {T}-gate codes of
  short length and high distance,'' \emph{IEEE J. Sel. Areas Inf. Theory},
  vol.~6, pp. 127--137, 2025.

\bibitem{BetsumiyaMunemasa12_1}
K.~Betsumiya and A.~Munemasa, ``On triply even binary codes,'' \emph{J. London
  Math. Soc.}, vol.~86, no.~1, pp. 1--16, Feb. 2012.

\bibitem{ShiLuKimSole24_1}
M.~Shi, H.~Lu, J.-L. Kim, and P.~Sol{\'e}, ``Triorthogonal codes and self-dual
  codes,'' \emph{Quantum Inf. Process.}, vol.~23, no.~7, Jul. 2024, {Art.} no.
  280.

\bibitem{Sullivan24_1}
M.~Sullivan, ``Code conversion with the quantum {Golay} code for a universal
  transversal gate set,'' \emph{Phys. Rev. A}, vol. 109, no.~4, Apr. 2024,
  {Art.} no. 042416.

\bibitem{KriegClausiusRuebenacketenBrink25_1}
F.~Krieg, J.~Clausius, M.~R{\"u}benacke, and S.~ten Brink, ``A comparative
  study of ensemble decoding methods for short length {LDPC} codes,'' in
  \emph{Proc. 14th Int. ITG Conf. Syst., Commun. Coding (SCC)}, Karlsruhe,
  Germany, Mar. 10--13, 2025, pp. 1--6.

\bibitem{MandelbaumBeznerJakeltenBrinkSchmalen26_1sub}
J.~Mandelbaum, P.~Bezner, H.~J{\"a}kel, S.~ten Brink, and L.~Schmalen, ``Affine
  subcode ensemble decoding of linear block codes,'' Apr. 2026,
  arXiv:2604.06889v1 [cs.IT].

\bibitem{KoutsioumpasSayginelWebsterBrowne25_1sub}
S.~Koutsioumpas, H.~Sayginel, M.~Webster, and D.~E. Browne, ``Automorphism
  ensemble decoding of quantum {LDPC} codes,'' Mar. 2025, arXiv:2503.01738v1
  [quant-ph].

\bibitem{DuffyLiMedard19_1}
K.~R. Duffy, J.~Li, and M.~M{\'e}dard, ``Capacity-achieving guessing random
  additive noise decoding,'' \emph{IEEE Trans. Inf. Theory}, vol.~65, no.~7,
  pp. 4023--4040, Jul. 2019, code available at
  \url{https://github.com/kenrduffy/GRAND-MATLAB}.

\bibitem{CruzMonteiroCoutinho23_1}
D.~Cruz, F.~A. Monteiro, and B.~C. Coutinho, ``Quantum error correction via
  noise guessing decoding,'' \emph{{IEEE Access}}, vol.~11, pp.
  119\,446--119\,461, 2023.

\bibitem{CalderbankRainsShorSloane98_1}
A.~R. Calderbank, E.~M. Rains, P.~W. Shor, and N.~J.~A. Sloane, ``Quantum error
  correction via codes over {GF(4)},'' \emph{IEEE Trans. Inf. Theory}, vol.~44,
  no.~4, pp. 1369--1387, Jul. 1998.

\bibitem{MacWilliamsSloane77_1}
F.~J. MacWilliams and N.~J.~A. Sloane, \emph{{The Theory of Error-Correcting
  Codes}}.\hskip 1em plus 0.5em minus 0.4em\relax Amsterdam, The Netherlands:
  North-Holland, 1977.

\bibitem{KohnertWassermann09_1}
A.~Kohnert and A.~Wassermann, ``Construction of binary and ternary
  self-orthogonal linear codes,'' \emph{Discrete Appl. Math.}, vol. 157, no.~9,
  pp. 2118--2123, May 2009.

\bibitem{HuffmanKimSole21_1}
W.~C. Huffman, J.-L. Kim, and P.~Sol{\'e}, Eds., \emph{{Concise Encyclopedia of
  Coding Theory}}.\hskip 1em plus 0.5em minus 0.4em\relax Boca Raton, FL, USA:
  CRC Press, 2021.

\bibitem{BravyiCross15_1sub}
S.~Bravyi and A.~Cross, ``Doubled color codes,'' Sep. 2015, arXiv:1509.03239v1
  [quant-ph].

\bibitem{RoffeWhiteBurtonCampbell20_1}
J.~Roffe, D.~R. White, S.~Burton, and E.~Campbell, ``Decoding across the
  quantum low-density parity-check code landscape,'' \emph{Phys. Rev. Res.},
  vol.~2, no.~4, Dec. 2020, {Art.} no. 043423.

\end{thebibliography}

\end{document}